\documentclass{aastex63}

\usepackage{graphicx}
\usepackage{amsmath}
\usepackage[utf8]{inputenc} 
\usepackage[T1]{fontenc}    
\usepackage{url}            
\usepackage{booktabs}       
\usepackage{amsfonts}       
\usepackage{nicefrac}       
\usepackage{microtype}      
\usepackage{lipsum}
\usepackage{listings}
\usepackage[frozencache,cachedir=.]{minted}
\usepackage{fancyvrb}

\newcommand{\trex}[1]{TauREx~#1}
\newcommand{\taurex}{TauREx}

\setminted{breaklines=true}
\setminted{breakbytokenanywhere=true}
\shorttitle{TauREx 3.1}
\shortauthors{Al-Refaie et al.}

\begin{document}

\title{A comparison of chemical models of exoplanet atmospheres enabled by TauREx 3.1}

\author{A. F. Al-Refaie}
\affiliation{Dept. Physics \& Astronomy, University College London, Gower Street, WC1E 6BT, London, UK}

\correspondingauthor{A. F. Al-Refaie}
\email{ahmed.al-refaie.12@ucl.ac.uk}

\author{Q. Changeat}
\affiliation{Dept. Physics \& Astronomy, University College London, Gower Street, WC1E 6BT, London, UK}

\author{O. Venot}
\affiliation{Université de Paris and Univ Paris Est Creteil, CNRS, LISA, F-75013 Paris, France}

\author{I. P. Waldmann}
\affiliation{Dept. Physics \& Astronomy, University College London, Gower Street, WC1E 6BT, London, UK}

\author{G. Tinetti}
\affiliation{Dept. Physics \& Astronomy, University College London, Gower Street, WC1E 6BT, London, UK}

\begin{abstract}
Thermochemical equilibrium is one of the most commonly used assumptions in current exoplanet retrievals. As the James Webb Space Telescope (JWST) and Ariel launch draw near, assessing the underlying biases and assumptions made when applying self-consistent chemistry into spectral retrievals is crucial. Here we use the flexibility of TauREx 3.1 to  cross-compare  three state of the art chemical equilibrium codes: ACE, FastChem and GGchem. We simulate JWST spectra for ACE, FastChem, GGchem and GGchem+condensation containing only C, H, O, N elements and spectra for FastChem, GGchem and GGchem+condensation with a more extensive range of elements giving seven simulated JWST spectra in total and then cross-retrieve giving a total of 49 retrievals. 
\newline
Our analysis demonstrates that like-for-like, all chemical codes retrieve the correct parameters to $<1\%$ of the truth. However, retrievals, where the contained elements do not match the truth, parameters such as metallicity deviate by $20\%$ while maintaining extremely low uncertainties $<1\%$ giving false confidence. This point is of major importance for future analyses on JWST and Ariel, highlighting that self-consistent chemical scheme that do not employ the proper assumptions (missing species, fixed elemental ratios, condensation) are at risk of confidently biasing interpretations. Free chemistry retrievals employing parametric descriptions of the chemical profiles can provide alternative unbiased explorations.

\end{abstract}

%

\section{Introduction}

In the past fifteen years an increasing number of exoplanetary atmospheres have been observed using space and ground-based facilities \citep{Sing_2016, Tsiaras_2018, Welbanks_2019, Pinhas_2019}. Current data are often sparse and with low signal to noise, leading to degeneracies that require the use of sophisticated modelling tools to be interpreted correctly. Despite these limitations, the atmospheres of large ultra hot-Jupiters \citep{Haynes_2015_w33, Evans_2019, Edwards_2020_ares1, Evans_2020, Gandhi_2020,Changeat_2021_k9}, hot-Jupiters \citep{Tinetti_2007, Swain_2008, Swain_2009, Swain_2009_2, Line_2014, Stevenson_2014, Stevenson_2017, MacDonald_hd209, Changeat_2020_Kelt11, Pluriel_2020_Ares3, Changeat_2021_phasecurve} and hot-Saturn \citep{Nikolov_2018, Skaf_2020_Ares2, Anisman_2020,Yip_2021_gb} are regularly characterised using the transit, eclipse and phase-curve techniques. Smaller transitional planets are notoriously more challenging, but the Hubble Space Telescope has allowed us to infer some constraints for a few of these worlds \citep{Kreidberg_2014, DeWit_2016, Tsiaras_2016_55cnc, DeWit_2018, Tsiaras_2019_k218, Benneke_2019_k218, madhu_2020_k218, Guilluy_2021_Ares3, Edwards_2021_lhs, Mugnai_2021, Swain_2021}.
Successful characterization of atmospheres must take into account many processes outside of radiative transfer. Incorporating phenomena such as chemical reactions, fluid dynamics, clouds and quantum effects is paramount to understand these extrasolar objects.  An influx of spectral retrieval models that solve the inverse problem has left modellers spoilt for choice.  Retrieval codes include: TauRex 3 \citep{alrefaie2019taurex}, NEMESIS \citep{Irwin_nemesis}, CHIMERA \citep{Line_chimera}, ARCiS \citep{Ormel_arcis, Min_2020_arcis}, BART \citep{harrington_bart}, petitRADTRANS \citep{Mollire_petitrad}, HELIOS \citep{Kitzmann, Lavie_helios}, POSEIDON \citep{MacDonald_hd209},  HyDRA \citep{Gandhi_retrieval}, SCARLET \citep{benneke_retrieval}, PLATON II \citep{Zhang_platon} and Pyrat-Bay \citep{cubillos_pirat}. Each retrieval code attempts to infer the abundance profiles of the atmospheric species. Sampling molecular abundances at each layer of the atmosphere exponentially inflates the sampling dimension to hundreds of parameters for each molecule. Pursuing this method of abundance retrieval will be riddled with parameter degeneracy and massive computation time. To address this issue, the molecular abundances are often parameterized either through heuristic or self-consistent means.

Heuristic parameterization extracts abundance profiles directly from the data. These types of parameterizations usually make straightforward assumptions about the chemistry in the atmosphere. An example is isoabundance profiles, where the molecular mixing ratios are invariant for all atmospheric depths. Profiles such as isoabundance help study the limits of what species can be identified and constrain their abundances based on the shape and uncertainty of the input spectrum. They are also advantageous as they represent species using a few parameters reducing the sampling space. Furthermore, they are fast computationally.
The dimensionality of the problem generally grows linearly with the number of molecules. 

Self-consistent parameterization relies on an underlying chemical model to produce molecular abundances. Such models significantly reduce the number of parameters required to generate abundance profiles and have a physical basis behind them. These models can be slower than heuristic, requiring some physical process simulation to generate the final chemical profile. The majority of these self-consistent models follow the same basic principle, i.e. they set the initial abundance of elements in the atmosphere and iterate until convergence. The approach taken by each model can differ significantly, including the thermochemical data used, iteration scheme and the chemical network employed.

One of the most commonly used self-consistent chemical models is the \textit{thermochemical equilibrium} approximation. This approximation assumes chemical reaction timescales are faster than the dynamical timescale and ignores photo-chemical processes. Chemical reaction rates are generally proportional to temperature and pressure, so the approximation holds in high temperature, high-pressure environments
such as those found in stars and ultra hot-Jupiters. For non-equilibrium or `disequilibrium' environments, the approximation is often used as the initial condition to facilitate faster convergence of chemical kinetics.
Each retrieval code attempts to solve the problem differently: e.g., HELIOS exploits GPU acceleration and FastChem \citep{fastchem} chemistry for self-consistent retrievals using Multinest \citep{multinest, buchner_pymultinest}. PLATON utilizes interpolation of precomputed GGchem \citep{ggchem} chemical tables to speed up retrievals using dynesty. petitRADTRANS takes a `no-frills' approach and only solves the radiative transfer problem, leaving it up to the user to provide the chemistry and sampler. TauREx 3 includes ACE \citep{Agundez2012, Agundez20} as the chemical equilibrium model but allowed a degree of flexibility in incorporating new chemical codes, forward models, samplers and retrieval parameters.

In its most general form, computing the chemical composition of an atmosphere in equilibrium requires minimizing the Gibbs free energy of the system given an initial atomic abundance. However, the nonlinear dependence of the Gibbs free energy on the number density of species in the system and linear constraints such as mass and charge conservation makes the computation highly demanding and non-trivial.
Various algorithms have been developed to solve for chemical equilibrium. In particular, exploiting the law of mass action \citep{massaction} and thermochemical equilibrium constants $K$ transforms the problem into a series of $N$ unknowns by eliminating molecular number densities. The system is solvable using a root-finding algorithm such as Newton-Raphson method. High temperatures ($T\geq 1000~K$) coverage quickly and are numerically stable, but lower temperatures
become more untenable as $K > 10^{1000}$ which cannot be represented in standard double-precision floating-point. 

Equilibrium constants $K$ for species $i$ can be derived from the Gibbs free energy of formation:
\begin{equation}
    \ln K_i = -\frac{\Delta {G}_{f}^\circ}{RT}
\end{equation}
where $\Delta {G}_{f}^\circ$ is the Gibbs free energy of formation, $R$ is the universal gas constant, and $T$ is the temperature.
Sources of the $\ln K_i$ include the NIST-JANAF \citep{chaseJANAF} database but must be fitted as a function of temperature before they can be used. Various fitting functions have been proposed in the literature including the form by \cite{Tsuji73}:
\begin{equation}\label{eq:tsuji_form}
    \ln K_i = a_0 + a_1\theta + a_2\theta^2 + a_3\theta^3 + a_4\theta^4
\end{equation}
where $a_n$ are fitting coefficients and $\theta=\frac{5040}{T}$ where $T$ is temperature in Kelvin. Another form used are the NASA polynomials formalism given
by \cite{mcbride93}:


\begin{align} 
  H_i     &= a_0 + a_1\frac{T}{2} + a_2\frac{T^2}{3} + a_3\frac{T^3}{4}+ a_4\frac{T^4}{5} + \frac{a_5}{T} \nonumber \\
  S_i     &= a_0\ln T + a_1T + a_2\frac{T^2}{2} + a_3\frac{T^3}{3}+ a_4\frac{T^4}{4} + a_6 \nonumber \\
  \ln K_i &= H_i - S_i \label{eq:nasa_form}
\end{align}

where $H_i$ and $S_i$ are the enthalpy and entropy for species $i$ respectively. 

To summarise, building a chemical equilibrium code will require three choices: (1) the choice in an algorithm for minimizing the Gibbs free energy. In addition, if low-temperature applicability is required, the algorithm must also deal with numerical stability issues. (2) The source of the equilibrium constants $K$ and (3) the choice in fitting function for $\ln K$.

The underlying biases associated with the Gibbs minimization strategy, thermochemical data source, fitting function and choice of elements and their initial abundances are also not well known and may significantly affect the direction of an atmospheric retrieval. Studying the effects of each chemical code on atmospheric retrieval is non-trivial. Each retrieval code is essentially married to a particular chemical model. Switching between different retrieval codes can introduce variability from the radiative transfer implementation (such as k-table and cross-section opacities) and the choice of Bayesian sampler. 
\trex{3.1} (Appendix \ref{ap:t31}) remedies this issue through the introduction of a plugin system. This system allows for light packages
that can be built, distributed through PyPI and installed. These packages are then detected by \taurex\ and seamlessly `plugged in' to the framework (see Appendix \ref{ap:plugins} for more details). Plugins can extend and enhance every aspect of the
framework, including but not limited to: GPU-acceleration (Appendix \ref{sec:cuda}), forward models, opacities, prior functions, samplers  and importantly, chemical models (Appendix \ref{ap:chemistry}).

We exploit \trex{3.1} and its plugin feature to compare and cross-validate three chemical codes: ACE \citep{Agundez2012, Agundez20}, Fastchem \citep{fastchem} and GGchem \citep{ggchem} under the \taurex\ retrieval framework. 

\section{METHODS}

\subsection{Description of chemical codes}

\subsubsection{ACE}
The ACE chemical FORTRAN code \citep{Agundez2012, Agundez20} was developed with a primary focus on studying chemical compositions in exoplanet atmospheres and is the equilibrium model in the Ariel target retrieval study by \cite{alfnoor}.
ACE minimizes the Gibbs free energy using the algorithm outlined by \cite{gordon94} and the included thermochemical data contains 105 neutral C/H/O/N bearing chemical species sourced from \cite{gordon94}, \cite{Atkinson06}, \cite{Bounaceur10} and with some constants computed using THERGAS \citep{Muller95}. The equilibrium constants are fit with the NASA polynomial formalism (Eq. \ref{eq:nasa_form}) applicable to the 300--5000~K temperature range. 

The ACE code supports a range of species and ions but its provided thermochemical data is heavily focused towards neutral C/H/O/N species as it is commonly used in conjunction with chemical kinetic models by \cite{venotdiseq} and \cite{venotmeth2020}. Many of the reactions and species in the respective networks derive a portion of their rates from combustion mechanisms which deal with C, H, N and O bearing molecules in temperatures up to 3000~K and pressures up to 100 bar. Such conditions are similar to those found in Ultra Hot Jupiters; ACE provides almost double the number of C/H/O/N bearing species compared to other thermochemical datasets such as \cite{chaseJANAF}.

\subsubsection{Fastchem}

FastChem \citep{fastchem} is a C++ chemical equilibrium code from the Exoclimes simulation platform that implements a new semi-analytical method to minimize the Gibbs energy. It decomposes
the system into a set of singular variable coupled nonlinear equations. FastChem also exploits quadruple precision and the Nelder-Mead optimization algorithm to converge quickly at temperatures as low as 100~K. Thermochemical data for the 396 neutral and 114 charge species is sourced from the NIST-JANAF database \citep{chaseJANAF} and fit the expression:
\begin{equation} \label{eq:stock}
    \ln K = \frac{a_0}{T} + a_1\ln T + a_2 + a_3T + a_4T^2
\end{equation}
FastChem is the main chemical model used in the HELIOS \citep{Kitzmann, Lavie_helios} radiative transfer code and has seen successful use in retrievals
of brown dwarf companions HD 1160B and HD 19467B \citep{Mesa20} and secondary eclipses of WASP-121b \citep{Bourrier20}

\subsubsection{GGchem}

GGchem (Gleich-Gewichts-Chemie) \citep{ggchem} is a FORTRAN-90 rewrite of the code that was originally used to study condensation processes in late M-type stars. The rewrite updates the iteration scheme extends numerical accuracy to quadruple precision, improves the range of applicability down to 100 K, and includes equilibrium condensation. GGchem supports the elements from hydrogen to zirconium plus tungsten and includes ion chemistry. GGchem supports the widest range of fitting function which include forms given by \cite{Gail86}, \cite{Sharp90} and Equations \ref{eq:tsuji_form}, \ref{eq:nasa_form} and \ref{eq:stock}.
GGchem can exploit multiple sources of thermochemical data. GGchem will continuously add new species from each file; any species already defined will be overwritten with the newer data. By default, GGchem will load $\ln K$ for diatomics and atoms from \cite{Barklem16} fit to Eq. \ref{eq:stock}, NIST-JANAF \citep{chaseJANAF} fit to Eq. \ref{eq:stock} and some additional refits of \cite{Tsuji73} fit to Eq. \ref{eq:stock} giving in total 448 neutral and 117 charge species. GGchem is extensive in the atmospheric retrieval literature; it was recently used to analyze the atmospheres of the Hot Jupiters KELT-9b \citep{Kelt-9b} and also WASP-103b \citep{wasp103b}.

\subsubsection{Free chemistry}

In this paper, we also compare our results with the three chemical equilibrium codes to the baseline \trex{3.1} free chemistry. The free chemistry assumption is a conservative approach as it is only informed by the spectral features in the observed spectrum and does not assume any particular physical or chemical processes. While the free chemistry module in \trex{3.1} allows for any parametric law to  describe the chemical profiles of each species, we only consider the  simplest assumption of constant profiles with altitude. In \cite{Changeat19layer}, a detailed example of more complex profile is presented.

\subsection{Forward Model Comparison}
\label{sec:fm_compare}

For each chemical equilibrium code we model a HD-209458\,b like planet as our test system with stellar and planetary parameters from \cite{Stassun17} given in Table \ref{tab:hd209-test}. The atmosphere is modelled  as a cloud-free isothermal transit spectrum that includes molecular absorptions, collisionally induced absorptions (CIA) from H$_2$-H$_2$ and H$_2$-He and Rayleigh scattering calculated for all molecules given by \cite{cox_allen_rayleigh}. We then simulate observed spectra as recorded with  the JWST NIRISS instrument with GR700XD grism and the NIRSPEC instrument with an G395M/F290LP filter. We generated the uncertainties using \texttt{taurex\_jwst} plugin, which uses the Exowebb library as its noise model backend. 

\begin{table}[ht]

\centering
\begin{tabular}{lr}
\hline\hline
Parameter & Value \\
\hline 
HD 209458 &  \\
\hline
$R_s$ & 1.19 $R_{\odot}$\\
$T_s$ & 6091.0 K \\
$K_{mag}$ & 6.308 \\
$D_s$ & 48.37 pc \\
$Z_s$ & 0.01 $Z_{\odot}$ \\
$M_s$ & 1.23 $M_{\odot}$ \\
\hline
HD 209458 b &\\

\hline
$R$ & 1.39 $R_{J}$ \\
$M$ & 0.73 $M_{J}$ \\
Semi-major axis & 0.0486 AU \\
$t_{period}$ & 3.5247 days \\
$t_{transit}$ & 11037.18 s \\
$T$ & 1500 K \\
$Z$ & $Z_{\odot}$ \\
C/O & 0.5 \\
\hline
\end{tabular}
\caption{Planetary and parent star parameters from \cite{Stassun17} used to generate the simulated HD-209458\,b JWST transit spectra.}
\label{tab:hd209-test}
\end{table}

\begin{table*}[ht]
\centering
\begin{tabular}{lll}
\hline\hline
Opacities & Type &  Ref.  \\
\hline

H$_2$-H$_2$  & CIA & \cite{abel_h2-h2}, \cite{fletcher_h2-h2} \\
H$_2$-He   & CIA & \cite{abel_h2-he} \\
H$_2$O     & Abs. & \cite{barton_h2o}, \cite{polyansky_h2o} \\
CH$_4$     & Abs. & \cite{hill_xsec}, \cite{exomol_ch4} \\
CO       & Abs. & \cite{li_co_2015} \\
CO$_2$     & Abs. & \cite{rothman_hitremp_2010} \\
NH$_3$     & Abs. & \cite{Yurchenko_nh3} \\
K & Abs. & \cite{Allard16} \\
Na & Abs. & \cite{Allard19} \\
C$_2$H$_2$ & Abs. & \cite{Chubb20} \\
SO$_2$     & Abs. & \cite{Underwood16} \\
TiO     & Abs. & \cite{McKemmish19} \\
VO     & Abs. & \cite{McKemmish16} \\
\hline
\end{tabular}
\caption{List of opacities used}\label{tab:opacity-source}
\end{table*}

For Fastchem and GGchem, we model two cases: The \textit{C/H/O/N} case where we limit the available elements to H, He, C, N, and O to allow for direct comparison with the ACE chemistry model.  
The second case loosens the element limitation by including H, He, C, O, N, Mg, S, Ti, V, K, and Na into the chemistry models. We defined this case  as heavy chemistry. 
For GGchem, we also include a particular case for both C/H/O/N and heavy chemistries, where we introduce sequestration from condensation.
In total, there are seven forward models produced. 


\subsection{Retrieval Cross-Validation}
\label{sec:retrieval_compare}
For each of the simulated JWST transit spectra, we cross-retrieve all seven combinations of chemical equilibrium codes and element lists. We also retrieve each case using a free chemistry model which includes an isoabundance profile for H$_2$O, CH$_4$, NH$_3$, C$_2$H$_2$, CO, CO$_2$, TiO and VO. In total, we therefore compare the results of X retrievals.
For this exercice,  we do not scatter the spectra as it has been repeatedly demonstrated \citep{Feng18, Changeat19layer, alfnoor} that for normally distributed noise and a large number of observations,  scattered and unscattered spectra are equivalent due to the central limit theorem.

We will only focus on three cases:
C/H/O/N chemistry and Heavy chemistry models fit against a C/H/O/N input spectrum, Heavy chemistry and ACE fits against a heavy input spectrum and Heavy chemistry and ACE fit against an input spectrum using the heavy condensation model. The cases where the chemical code elemental composition and processes match the input spectrum will assess the underlying biases associated with each codes methodology, thermochemical data source and equilibrium constant function choice. Conversely, the cases where the elemental compositions and processes do not match will assess the biases associated with our initial assumption of the atmosphere's chemistry.
Our retrieval setup utilizes TauREx-CUDA contributions (Section \ref{sec:cuda}) to accelerate the forward modeling computation and the MultiNest Bayesian nested sampling code \citep{multinest, buchner_pymultinest} with Message Passing Interface (MPI). 
The nested sampling utilizes 750 live points, with an evidence tolerance of 0.5 and uniform priors given by Table \ref{tab:chem-priors}. For the free retrieval, we use priors given by Table \ref{tab:free-priors} and derive $Z$ and $C/O$ after the retrieval through post-processing \citep{Lee_2013, Macdonald_2019}. We perform the retrievals on the OzSTAR Supercomputing @ Swinbourne cluster; Each node has 2 Intel Gold 6140 18-core processors and two nVidia P100 12GB PCIe GPUs. 
For every retrieval, we use five nodes on the cluster. On each node, we only allocate four of the available cores and spawn four TauREx MPI processes. Each nVidia P100 is assigned two processes in order to maximize the GPU resources. In total, each retrieval is run on 20 cores exploiting 10 GPUs.

\begin{table}[ht]

\centering
\begin{tabular}{ll}
\hline\hline
Parameter & Prior bounds \\
\hline
$R_p$ & [0.5--2.0] $R_J$\\
$T$ &  [500--2000] K \\
$Z$ &  [0.1--4.0] $Z_{\odot}$\\
$C/O$ & [0.1--4.0] \\
\hline
\end{tabular}
\caption{Uniform priors used by ACE, FastChem and GGchem in the retrieval.}
\label{tab:chem-priors}
\end{table}

\begin{table}[ht]

\centering
\begin{tabular}{ll}
\hline\hline
Parameter & Prior bounds \\
\hline
$R_p$ & [0.5--2.0] $R_J$\\
$T$ &  [500--2000] K \\
$\log$ H$_2$O & [-12-- -2] \\
$\log$ CH$_4$ & [-12-- -2] \\
$\log$ NH$_3$ & [-12-- -2] \\
$\log$ C$_2$H$_2$ & [-12-- -2] \\
$\log$ CO & [-12-- -2] \\
$\log$ CO$_2$ & [-12-- -2] \\
$\log$ TiO & [-12-- -2] \\
$\log$ VO & [-12-- -2] \\

\hline
\end{tabular}
\caption{Uniform priors used by the free chemistry retrieval.}
\label{tab:free-priors}
\end{table}

Examining the computational performance of the retrievals, we required $\approx$ 40,000 samples for completion. Table \ref{tab:chemperf} displays the mean sampling times for each code. It is evident that the CUDA plugin greatly accelerates the retrieval compared to the previous version \citep{alrefaie2019taurex}, with retrieval times ranging from 4--30 minutes for most of the chemical codes and element list. The exception is the heavy condensation case, which required over 9 hours to complete. The calculation of equilibrium condensation on the CPU is a significant proportion of the forward modeling time; therefore, GPU acceleration of the optical depth is of little benefit in this particular case.

\begin{table}[ht]

\centering
\begin{tabular}{lll}
\hline\hline
Code & Species & Retrieval Time (min) \\
\hline
ACE & C/H/O/N & 4.40 \\
Fastchem & C/H/O/N & 4.45 \\
Fastchem & heavy & 10.55 \\
GGchem & C/H/O/N & 10.52\\
GGchem & heavy &  30.2 \\
GGchem & heavy+cond & 540.43\\
\hline
\end{tabular}
\caption{Average retrieval times for each chemical code and species used. Benchmarks were conducted on the OzSTAR Supercomputing @ Swinbourne cluster on 5 nodes; Each node has 2 Intel Gold 6140 18-core processors and two nVidia P100 12GB PCIe GPUs. }
\label{tab:chemperf}
\end{table}

For all the plots, we will maintain the same color scheme, i.e. blue for ACE, red for Fastchem, green for GGchem, and orange for GGchem with condensation. 


\section{RESULTS}

\subsection{Forward Model Comparison}
\begin{center}
\begin{figure}[ht]
    \includegraphics[width=1.0\columnwidth]{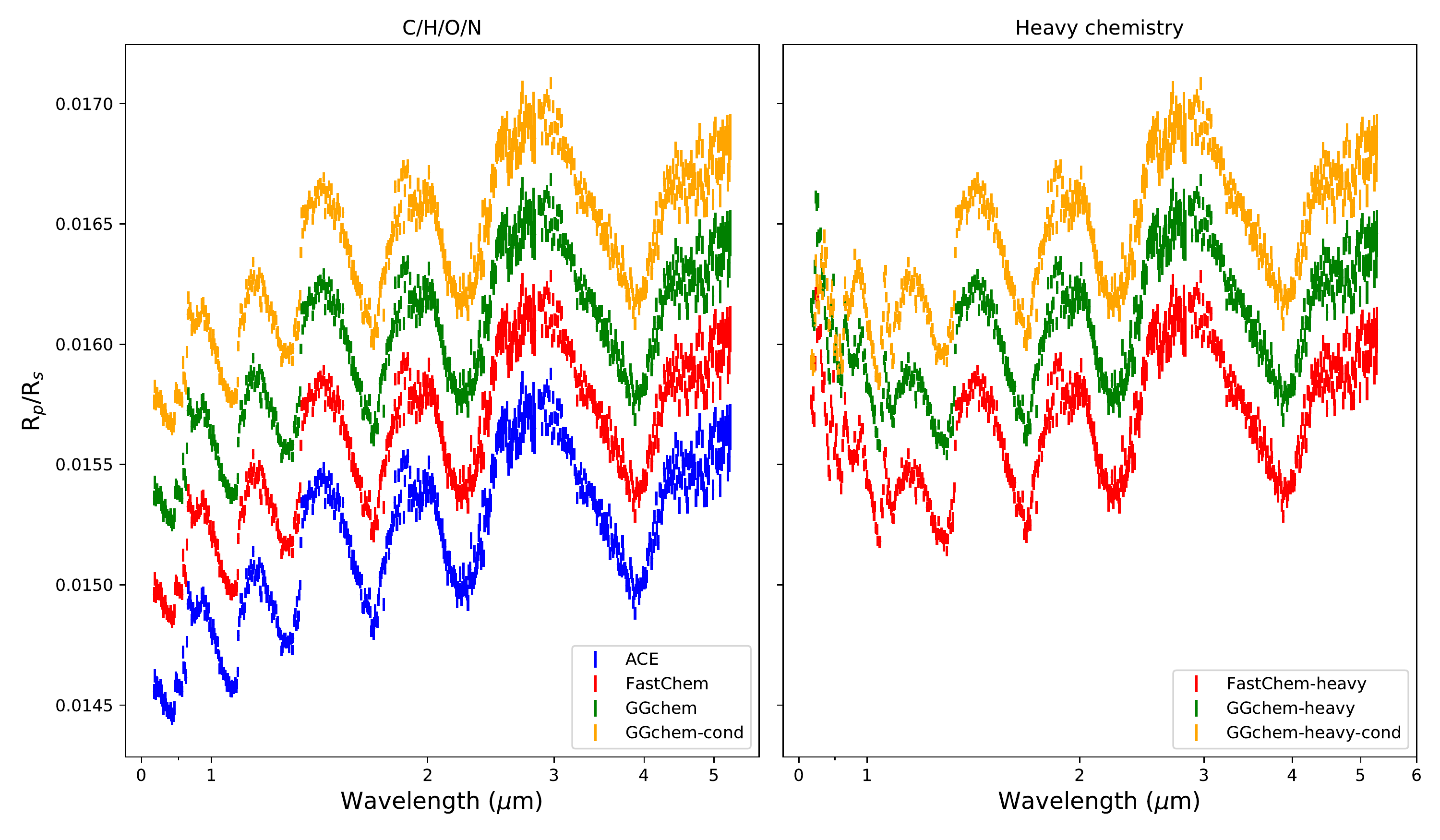}
\caption{Plots of simulated JWST transit spectra from different chemistry models for the test planet given in Table \ref{tab:hd209-test}. Each spectra above ACE has been artificially offset for clarity. \textit{C/H/O/N} chemistry restricts each equilibrium model to the elements H, He, C, N, and O. \textit{Heavy chemistry} includes H, He, C, O, N, Mg,
S, K, and Na and is suffixed with \textit{heavy}. The \textit{cond} suffix includes
condensation.}
\label{fig:allchem}
\end{figure}
\end{center}

Figure \ref{fig:allchem} shows simulated JWST  transit spectra obtained from different chemistry models for the test planet given in Table \ref{tab:hd209-test}.
Qualitatively the C/H/O/N case (Left panel of Fig. \ref{fig:allchem}) presents highly similar spectral features since water absorption dominates most of the spectrum. There is no condensation apparent when selecting only C/H/O/N species. Assessing the heavy case (Right panel of Fig. \ref{fig:allchem}), we observe strong absorption features from TiO (0.3~\micron--1~\micron) and an absorption feature from VO at 1.1~\micron\ in FastChem and GGchem that is not accounted for in ACE. 

\begin{center}
\begin{figure}[ht]
    \includegraphics[width=1.0\columnwidth]{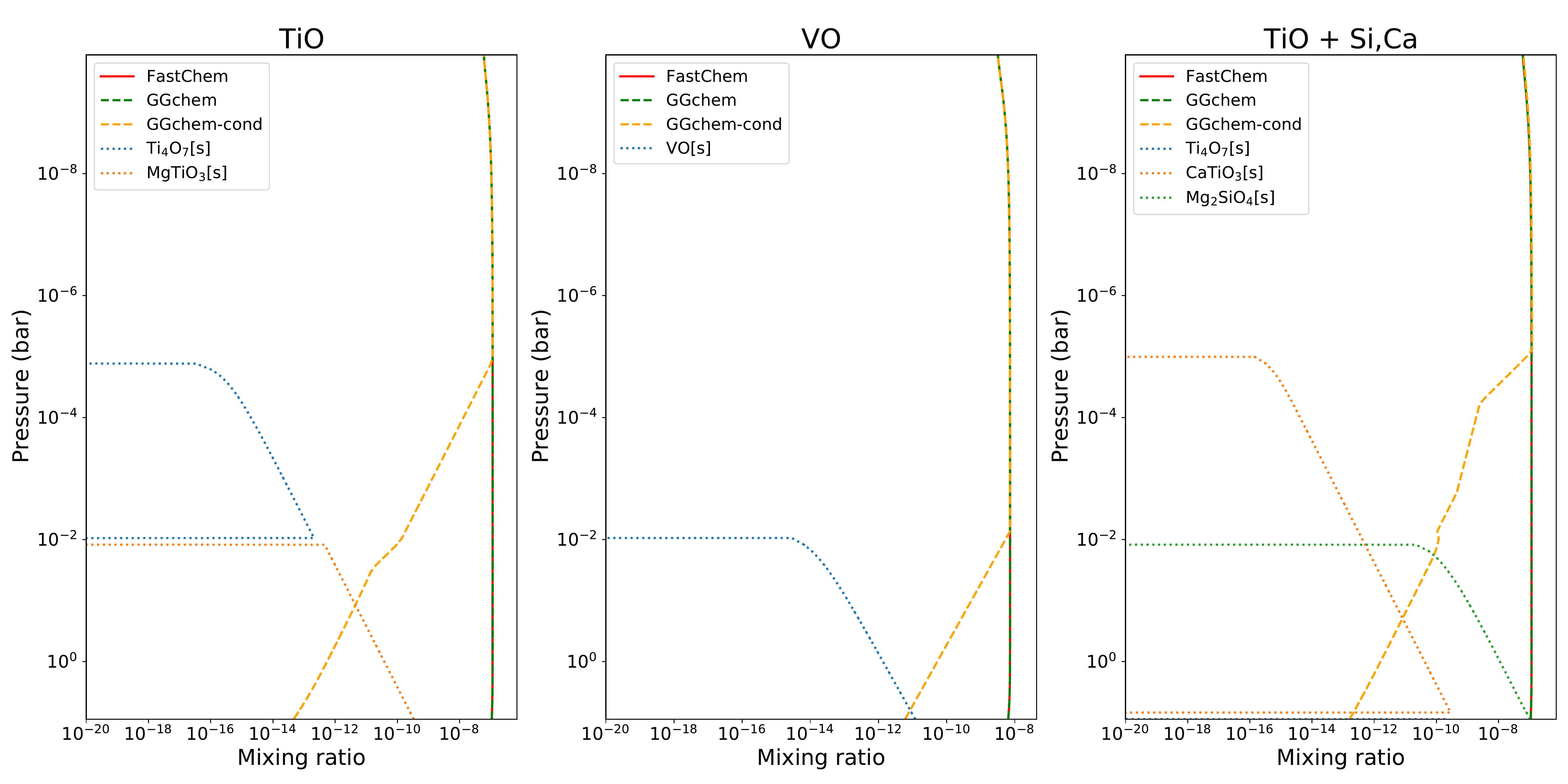}
\caption{Plots of volume mixing ratios of TiO and VO using the heavy chemistry for both FastChem, GGchem and GGchem with condensation. 
The rightmost plot is the heavy chemistry where silicon and calcium are included.}
\label{fig:TVcond}
\end{figure}
\end{center}

For the case where GGchem includes condensation, we observe a reduction in the spectral features in TiO and a minor reduction in VO which are due to the decrease of their abundance in the atmosphere. Figure \ref{fig:TVcond} shows clearly that these molecules condense in the low atmosphere. The main contributor to TiO loss comes from the formation of solid-phase MgTiO$_3$ in the troposphere, and Ti$_4$O$_7$ in the stratosphere (see e.g.,  \cite{Lodders_2002}). However, the overabundance of MgTiO$_3$ is a consequence of our choice of elements lacking silicon, which sequesters magnesium through forsterite (Mg$_2$SiO$_4$) seen in Figure \ref{fig:TVcond}. Additionally, the calcium-containing condensate CaTiO$_2$ is stable at the atmospheric region below 10$^{-5}$ bar and is the principal constituent of Ti-bearing condensates. CaTiO$_2$ is more efficient at removing TiO from the troposphere and hinders the production of Ti$_4$O$_7$. The reduction in Ti$_4$O$_7$ allows for a higher concentration of TiO in the stratosphere. 

For vanadium, the first and most stable condensate is vanadium monoxide in the solid-phase \citep{Burrows_1999, Allard_2001} which removes gas-phase VO from the lower atmosphere. The introduction of calcium to form vanadium analogies to the calcium titanates introduces no new condensates as these are not stable \citep{Lodders_2002}.

Overall, all equilibrium chemical models arrive at very similar conclusions when presented with the same elements. Differences only appear based on the choice of the elemental composition and condensation of the atmosphere rather than chemical code choice.

\subsection{Retrieval Cross-Validation}

\begin{table*}[ht]
\label{tab:summary}
\begin{tabular}{lcccc}
Model            &  C/H/O/N & C/H/O/N + cond & Heavy & Heavy + cond \\
\hline\hline \\
Radius ($R_J$)\\
\hline \\
ACE & 1.39$^{+0.00}_{-0.00}$ & 1.39$^{+0.00}_{-0.00}$ & 1.42$^{+0.00}_{-0.00}$ & 1.41$^{+0.00}_{-0.00}$ \\
FastChem & 1.39$^{+0.00}_{-0.00}$ & 1.39$^{+0.00}_{-0.00}$ & 1.42$^{+0.00}_{-0.00}$ & 1.41$^{+0.00}_{-0.00}$ \\
GGchem & 1.39$^{+0.00}_{-0.00}$ & 1.39$^{+0.00}_{-0.00}$ & 1.42$^{+0.00}_{-0.00}$ & 1.41$^{+0.00}_{-0.00}$ \\
GGchem+cond & 1.39$^{+0.00}_{-0.00}$ & 1.39$^{+0.00}_{-0.00}$ & 1.42$^{+0.00}_{-0.00}$ & 1.41$^{+0.00}_{-0.00}$ \\
FastChem (Heavy) & 1.39$^{+0.00}_{-0.00}$ & 1.39$^{+0.00}_{-0.00}$ & 1.39$^{+0.00}_{-0.00}$ & 1.38$^{+0.00}_{-0.00}$ \\
GGchem (Heavy) & 1.39$^{+0.00}_{-0.00}$ & 1.39$^{+0.00}_{-0.00}$ & 1.39$^{+0.00}_{-0.00}$ & 1.38$^{+0.00}_{-0.00}$ \\
GGchem (Heavy) + cond & 1.39$^{+0.00}_{-0.00}$ & 1.39$^{+0.00}_{-0.00}$ & 1.39$^{+0.00}_{-0.00}$ & 1.39$^{+0.00}_{-0.00}$ \\
Free & 1.39$^{+0.00}_{-0.00}$ & 1.39$^{+0.00}_{-0.00}$ & 1.39$^{+0.00}_{-0.00}$ & 1.39$^{+0.00}_{-0.00}$ \\
\hline \\
Temperature (K)\\
\hline \\
ACE & 1498.57$^{+3.85}_{-4.18}$ & 1498.54$^{+3.53}_{-4.12}$ & 1167.21$^{+5.22}_{-5.18}$ & 1343.73$^{+4.40}_{-4.55}$ \\
FastChem & 1495.27$^{+4.26}_{-4.46}$ & 1495.29$^{+4.27}_{-4.30}$ & 1166.49$^{+5.24}_{-5.27}$ & 1341.80$^{+4.34}_{-4.51}$ \\
GGchem & 1496.28$^{+4.02}_{-4.21}$ & 1496.56$^{+3.85}_{-4.33}$ & 1166.47$^{+5.20}_{-5.03}$ & 1342.52$^{+4.49}_{-4.66}$ \\
GGchem+cond & 1496.50$^{+3.85}_{-4.19}$ & 1496.57$^{+3.86}_{-4.29}$ & 1166.70$^{+5.08}_{-5.48}$ & 1342.41$^{+4.73}_{-4.66}$ \\
FastChem (Heavy) & 1329.74$^{+2.56}_{-2.45}$ & 1329.93$^{+2.52}_{-2.44}$ & 1498.85$^{+4.81}_{-5.28}$ & 1450.00$^{+2.14}_{-2.83}$ \\
GGchem (Heavy) & 1329.70$^{+2.61}_{-2.55}$ & 1330.04$^{+2.44}_{-2.45}$ & 1499.06$^{+4.70}_{-5.31}$ & 1448.77$^{+4.03}_{-3.71}$ \\
GGchem (Heavy) + cond & 1373.13$^{+1.85}_{-2.05}$ & 1372.16$^{+2.07}_{-2.40}$ & 1620.19$^{+3.11}_{-2.96}$ & 1499.70$^{+2.25}_{-2.32}$ \\
Free & 1497.75$^{+4.05}_{-4.46}$ & 1497.85$^{+4.02}_{-4.75}$ & 1499.39$^{+4.84}_{-5.28}$ & 1507.68$^{+4.59}_{-4.45}$ \\
\hline \\
Metallicity ($Z_{\odot}$)\\
\hline \\
ACE & 0.99$^{+0.08}_{-0.08}$ & 0.99$^{+0.08}_{-0.07}$ & 0.27$^{+0.02}_{-0.02}$ & 0.28$^{+0.02}_{-0.02}$ \\
FastChem & 0.98$^{+0.08}_{-0.08}$ & 0.97$^{+0.07}_{-0.07}$ & 0.27$^{+0.02}_{-0.02}$ & 0.27$^{+0.02}_{-0.02}$ \\
GGchem & 0.98$^{+0.08}_{-0.07}$ & 0.98$^{+0.07}_{-0.07}$ & 0.27$^{+0.02}_{-0.02}$ & 0.28$^{+0.02}_{-0.02}$ \\
GGchem+cond & 0.98$^{+0.08}_{-0.07}$ & 0.98$^{+0.07}_{-0.07}$ & 0.27$^{+0.02}_{-0.02}$ & 0.28$^{+0.02}_{-0.02}$ \\
FastChem (Heavy) & 4.00$^{+0.00}_{-0.00}$ & 4.00$^{+0.00}_{-0.00}$ & 1.00$^{+0.08}_{-0.07}$ & 3.99$^{+0.01}_{-0.01}$ \\
GGchem (Heavy) & 4.00$^{+0.00}_{-0.00}$ & 4.00$^{+0.00}_{-0.00}$ & 1.01$^{+0.08}_{-0.08}$ & 3.99$^{+0.01}_{-0.01}$ \\
GGchem (Heavy) + cond & 2.10$^{+0.13}_{-0.13}$ & 2.05$^{+0.12}_{-0.12}$ & 0.35$^{+0.03}_{-0.02}$ & 0.98$^{+0.08}_{-0.07}$ \\
Free & 0.69$^{+0.15}_{-0.12}$ & 0.70$^{+0.16}_{-0.12}$ & 0.66$^{+0.17}_{-0.13}$ & 0.85$^{+0.20}_{-0.16}$ \\
\hline \\
C/O\\
\hline \\
ACE & 0.50$^{+0.02}_{-0.02}$ & 0.55$^{+0.02}_{-0.02}$ & 0.36$^{+0.02}_{-0.02}$ & 0.63$^{+0.02}_{-0.02}$ \\
FastChem & 0.50$^{+0.02}_{-0.02}$ & 0.55$^{+0.02}_{-0.02}$ & 0.36$^{+0.02}_{-0.02}$ & 0.62$^{+0.02}_{-0.02}$ \\
GGchem & 0.50$^{+0.02}_{-0.02}$ & 0.55$^{+0.02}_{-0.02}$ & 0.35$^{+0.02}_{-0.02}$ & 0.63$^{+0.02}_{-0.02}$ \\
GGchem+cond & 0.50$^{+0.02}_{-0.02}$ & 0.55$^{+0.02}_{-0.02}$ & 0.35$^{+0.02}_{-0.02}$ & 0.63$^{+0.02}_{-0.02}$ \\
FastChem (Heavy) & 0.10$^{+0.00}_{-0.00}$ & 0.10$^{+0.00}_{-0.00}$ & 0.50$^{+0.01}_{-0.01}$ & 0.13$^{+0.00}_{-0.00}$ \\
GGchem (Heavy) & 0.10$^{+0.00}_{-0.00}$ & 0.10$^{+0.00}_{-0.00}$ & 0.50$^{+0.01}_{-0.01}$ & 0.12$^{+0.01}_{-0.01}$ \\
GGchem (Heavy) + cond & 0.30$^{+0.02}_{-0.02}$ & 0.34$^{+0.02}_{-0.02}$ & 0.59$^{+0.01}_{-0.01}$ & 0.55$^{+0.02}_{-0.02}$ \\
Free & 0.51$^{+0.04}_{-0.04}$ & 0.56$^{+0.04}_{-0.04}$ & 0.50$^{+0.04}_{-0.04}$ & 0.55$^{+0.04}_{-0.04}$ \\
\hline \\
\end{tabular}
\caption{A summary of all retrieved values and uncertainties for each combination of simulated spectra and chemical model. \textit{C/H/O/N} and \textit{C/H/O/N + cond} are ACE and GGchem+condensation with only C, H, O and N elements and simulated as JWST spectra respectively. \textit{Heavy} and \textit{Heavy + cond} are Fastchem and GGchem+condensation with elements H, He, C, O, N, Mg, S, Ti, V, K, and Na  and simulated as JWST spectra respectively.}
\end{table*}


Table \ref{tab:summary} summarises all retrieved values and uncertainties for each combination of simulated spectra and chemical model.
We find the same posterior distributions for all parameters as expected by matching the initial chemical elements with the simulated spectra. The actual implementation of equilibrium chemistry and the thermochemical tables and fitting functions used have no significant effect on the retrieval, and all give consistent results. For example, retrieving against ACE, FastChem with C/H/O/N and GGChem with C/H/O/N spectra give the same posteriors, as expected and described in the previous section.  
This result is important because the means of solving thermochemical equilibrium introduces no visible bias in the retrieval.
No single code produces the 'best' retrieval. Its choice is entirely a user preference based on needs such as ease-of-use, speed, thermochemical completeness, applicability and whether to include condensation.

By contrast, when the model species and input spectrum species do not match, we find a significant deviation between true and retrieved values. We discuss these results more in details in the following sections.

Regarding the free chemistry case, our retrievals manage to recover the input abundances simulated from the chemical equilibrium models. In all cases, C/H/O/N, Heavy, Heavy (condensation), the input spectra are well fit and the true abundances are within the retrieved uncertainties. This provides confidence that a free approach can be used to interpret exoplanet spectra, while avoiding the main user-dependent choices implied in chemical equilibrium codes.

\subsubsection{C/H/O/N input spectrum vs C/H/O/N models}

\begin{center}
\begin{figure}[ht]
    \includegraphics[width=1.0\columnwidth]{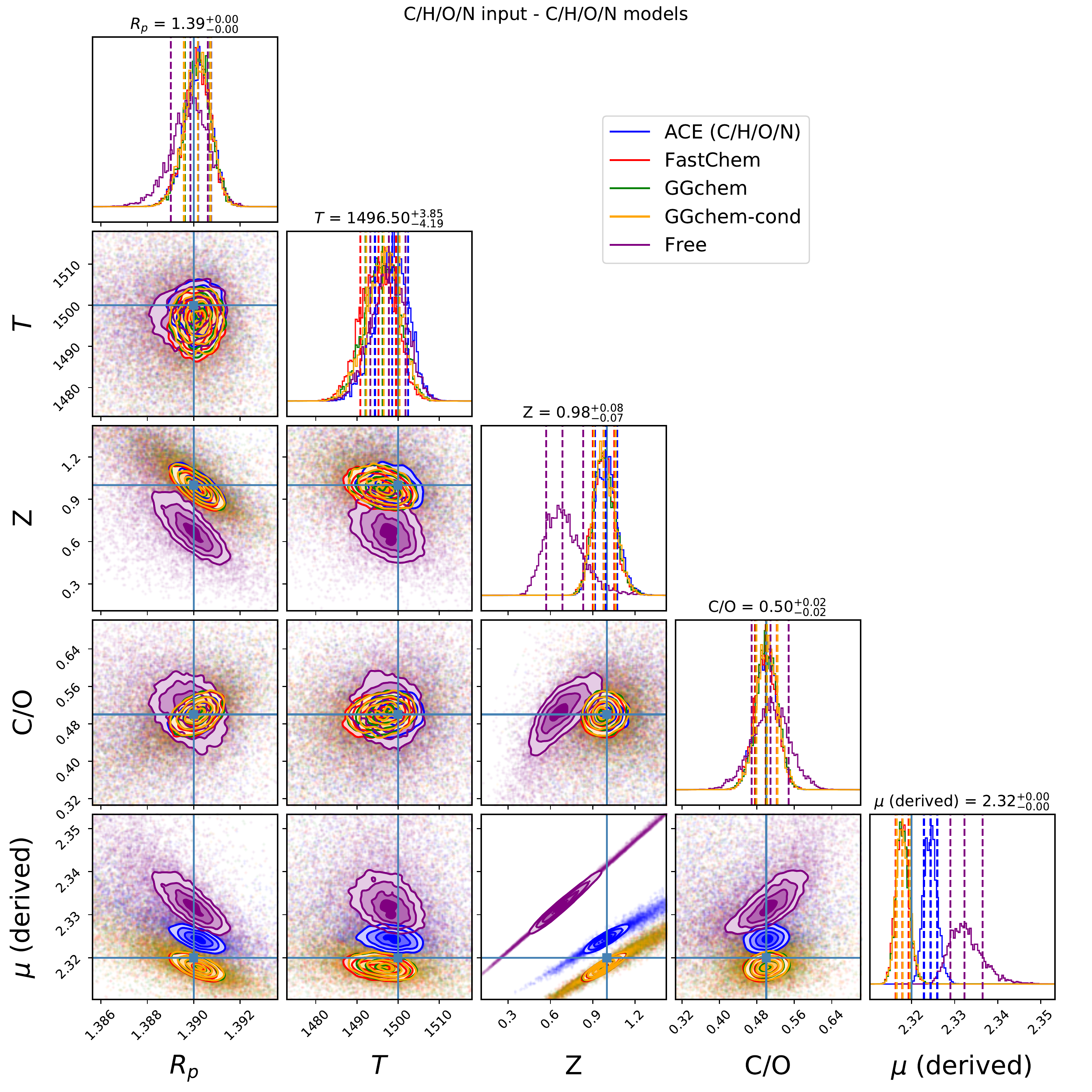}
\caption{Retrieval posteriors of ACE (blue), FastChem (red), GGChem (green), GGChem with condensation (orange) and free chemistry (purple). Each equilibrium chemistry code utilizes the C/H/O/N element list and is fit against a JWST simulated transit spectra of HD 209458b. Simulation (Truth) parameters are given by Table \ref{tab:hd209-test} and chemical abundances computed using ACE \citep{Agundez2012, Agundez20}. Priors used are given by Table \ref{tab:chem-priors}. }
\label{fig:chon-chon-post}
\end{figure}
\end{center}

\begin{center}
\begin{figure}[ht]
    \includegraphics[width=1.0\columnwidth]{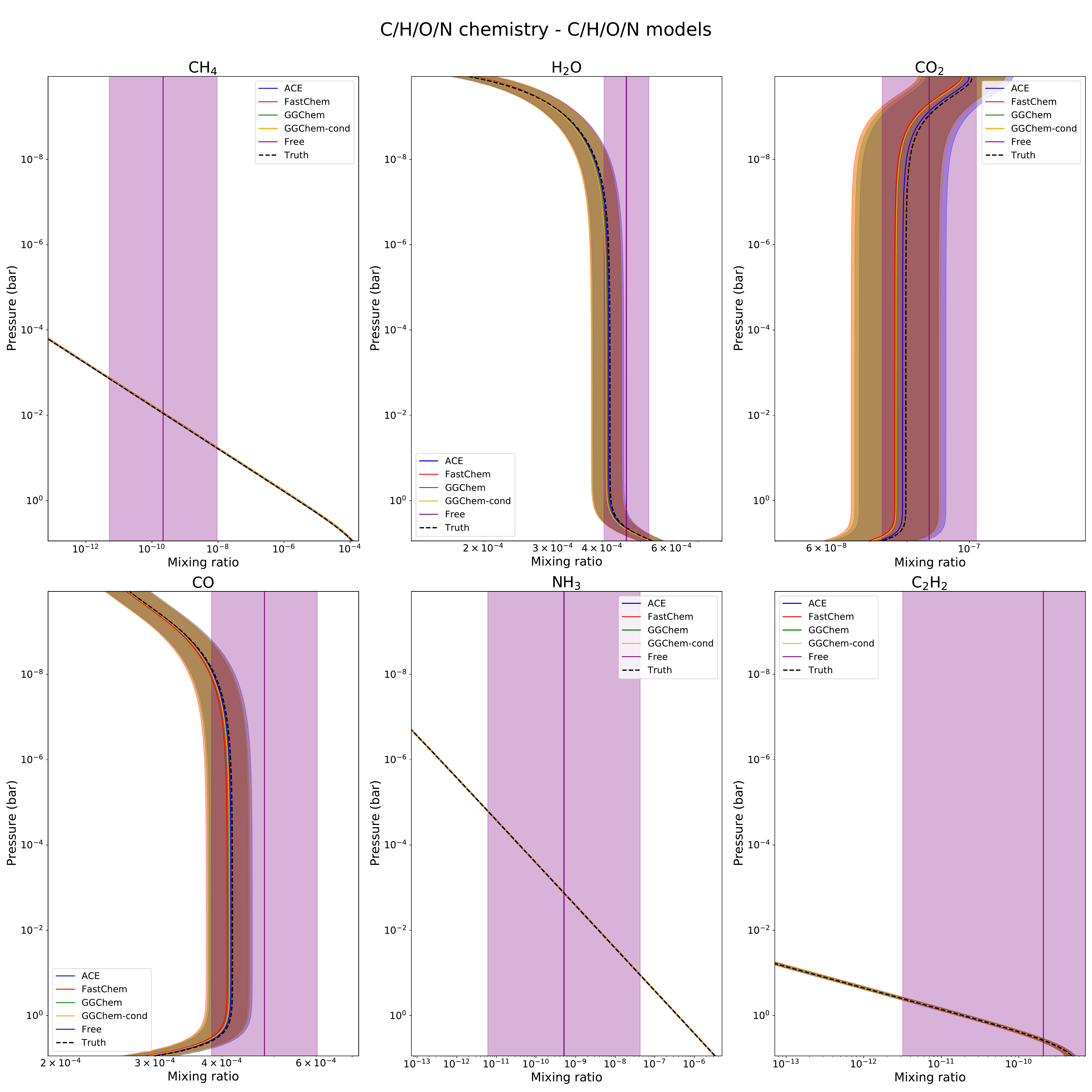}
\caption{Retrieved molecular profiles for ACE (blue), FastChem (red), GGChem (green), GGChem with condensation (orange) and free chemistry (purple). Each equilibrium chemistry code utilizes the C/H/O/N element list. Shaded regions denote 1$\sigma$ span, dashed lines are truth values computed from the ACE chemical code using Table \ref{tab:hd209-test}.}
\label{fig:chon-chon-prof}
\end{figure}
\end{center}

In Figure \ref{fig:chon-chon-post}, we compare C/H/O/N species for all codes against the ACE simulated spectra. The posteriors of Fastchem, GGchem, and GGchem with condensation are within 1$\sigma$ of ACE's posterior volume.  The mean molecular weight is slightly higher for ACE as it has almost double the number of molecules (105) compared to Fastchem (57) and GGchem (58) when selecting only C/H/O/N species. The retrieved atmospheric molecular profiles in Figure \ref{fig:chon-chon-prof} reveal that all codes recover well the correct molecular profile for all significant molecules present.

Looking at the retrieved chemical profiles, the free chemistry manages to recover the abundances of individual species very precisely. In turn, posteriors recover well  the radius, temperature, C/O ratio but underestimate the metallicity since many inactive molecules cannot be retrieved in the free chemistry model. 

\begin{center}
\begin{figure}[ht]
    \includegraphics[width=1.0\columnwidth]{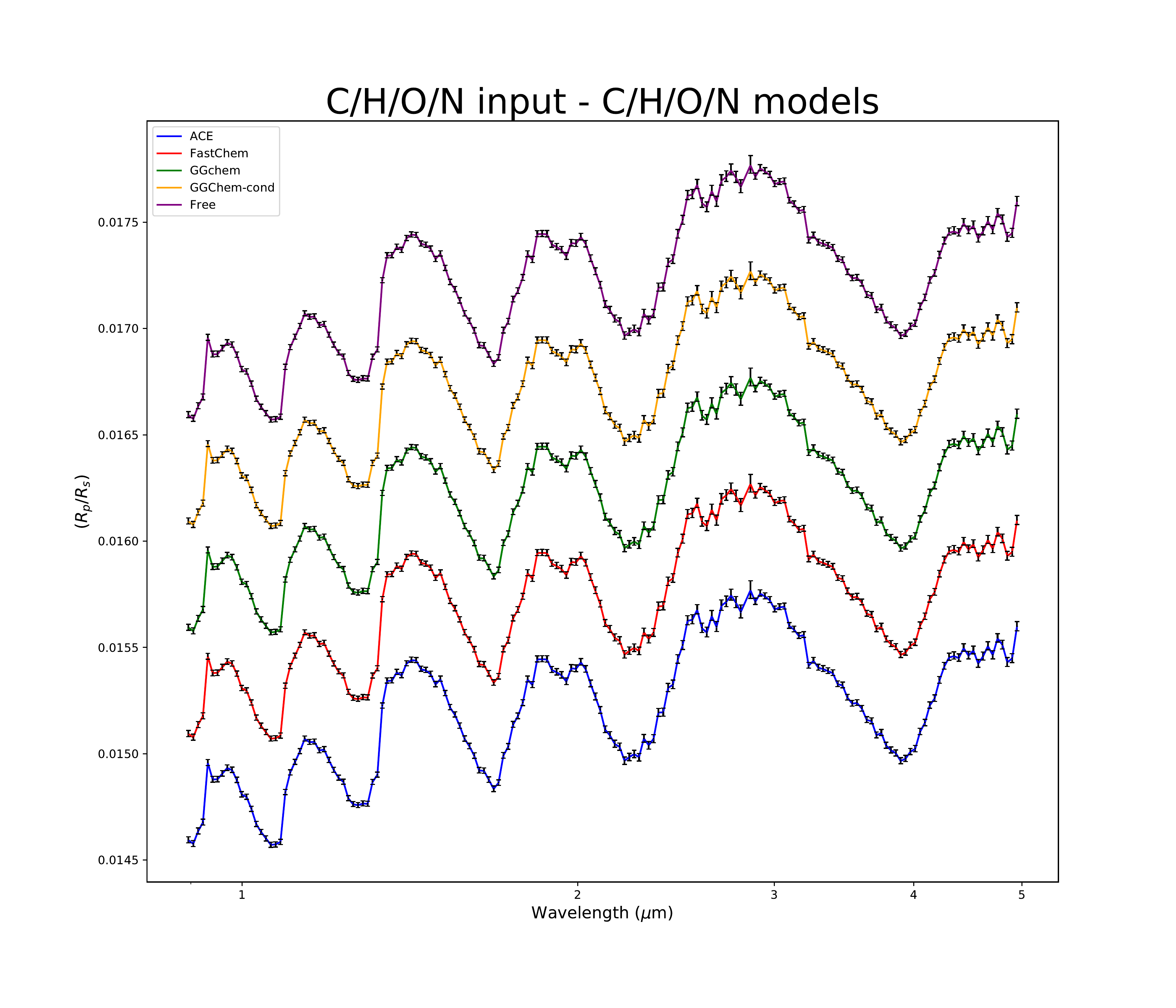}
\caption{Retrieved spectra for ACE (blue), FastChem (red), GGChem (green), GGChem with condensation (orange) and free chemistry (purple). Each equilibrium chemistry code utilizes the C/H/O/N element list and is retrieved against simulated JWST spectra (black) computed with ACE using parameters from Table \ref{tab:hd209-test}. Spectra have been binned to a lower resolution and offset for clarity.}
\label{fig:chon-chon-spec}
\end{figure}
\end{center}

Finally, the resultant spectra in Figure \ref{fig:chon-chon-spec} all lie within the uncertainties of the simulated JWST observations.

\subsubsection{C/H/O/N input spectrum vs Heavy models}

\begin{center}
\begin{figure}[ht]
    \includegraphics[width=1.0\columnwidth]{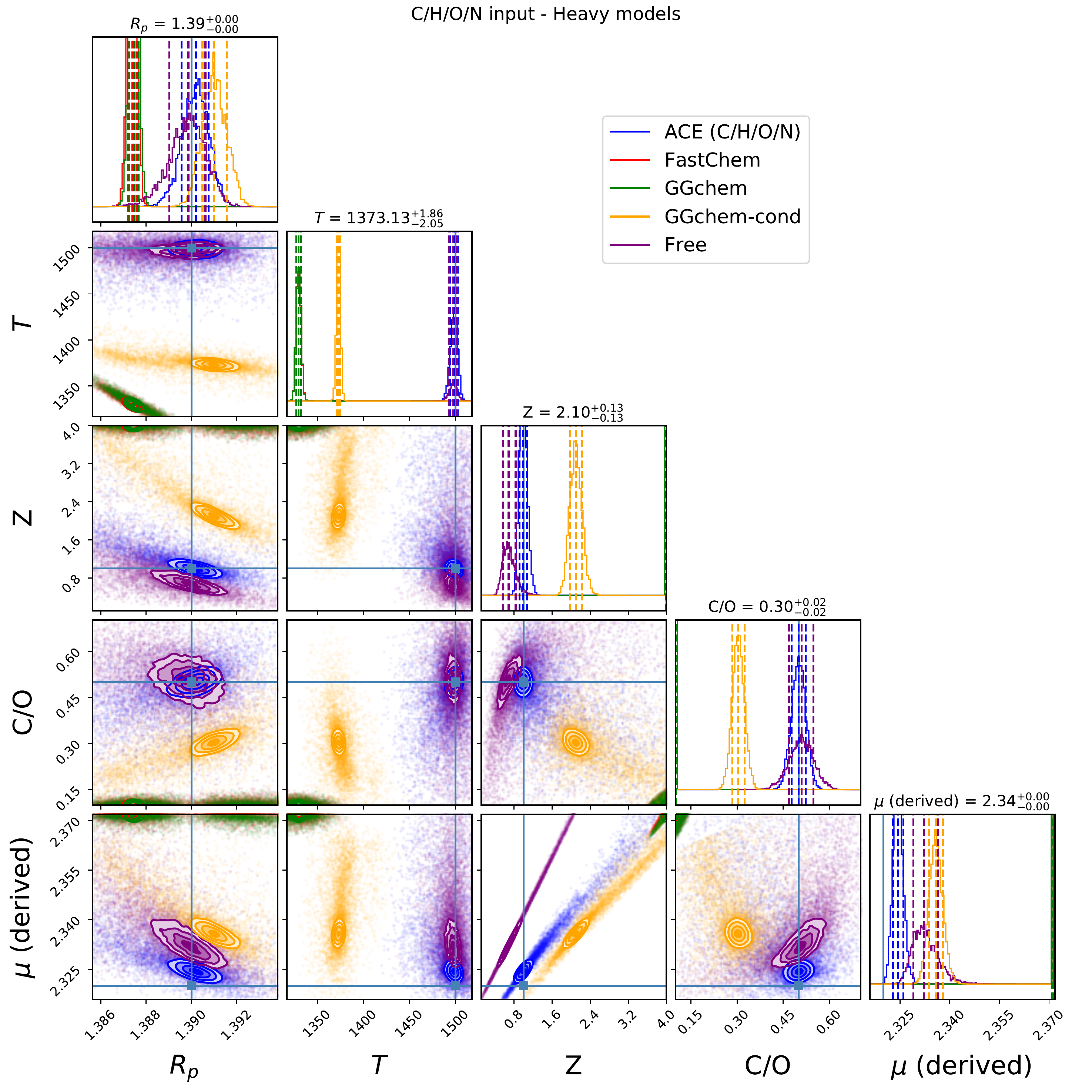}
\caption{Retrieval posteriors of ACE (blue), FastChem (red), GGChem (green) and GGChem with condensation (orange). Each equilibrium chemical code utilizes the heavy element list (except ACE and free) and is fit against a JWST simulated transit spectra of HD 209458b using parameters given by Table \ref{tab:hd209-test} and abundances computed using ACE \citep{Agundez2012, Agundez20}. Priors used are given by Table \ref{tab:chem-priors}.}
\label{fig:chon-heavy-post}
\end{figure}
\end{center}

\begin{center}
\begin{figure}[ht]
    \includegraphics[width=1.0\columnwidth]{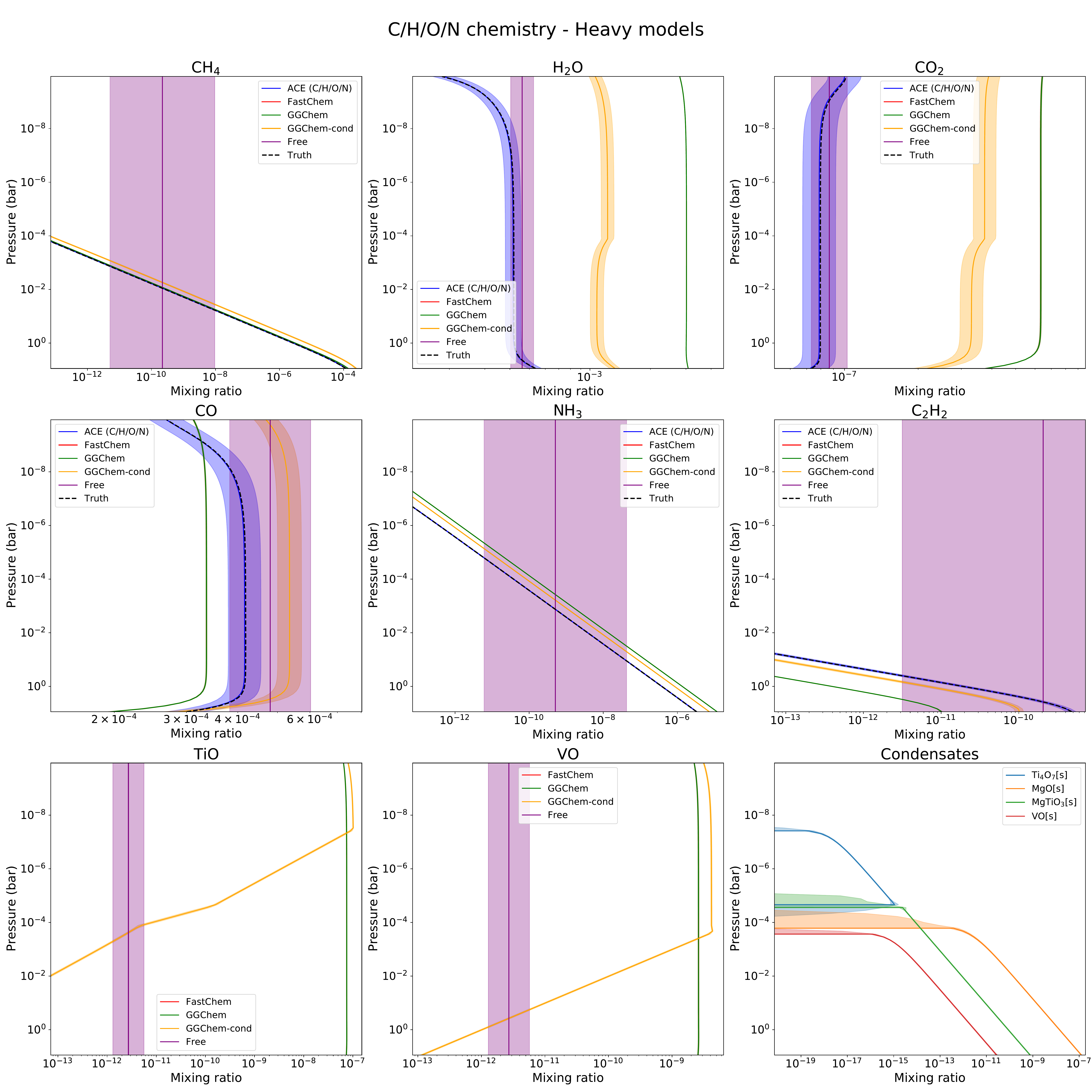}
\caption{Retrieved molecular profiles for ACE (blue), FastChem (red), GGChem (green), GGChem with condensation (orange) and free chemistry (purple). Each equilibrium chemical code utilizes the heavy element list (except ACE and free). Shaded regions denote 1$\sigma$ span, dashed lines are truth values computed from the ACE chemical code (C/H/O/N) using Table \ref{tab:hd209-test}.}
\label{fig:chon-heavy-prof}
\end{figure}
\end{center}

\begin{center}
\begin{figure}[ht]
    \includegraphics[width=1.0\columnwidth]{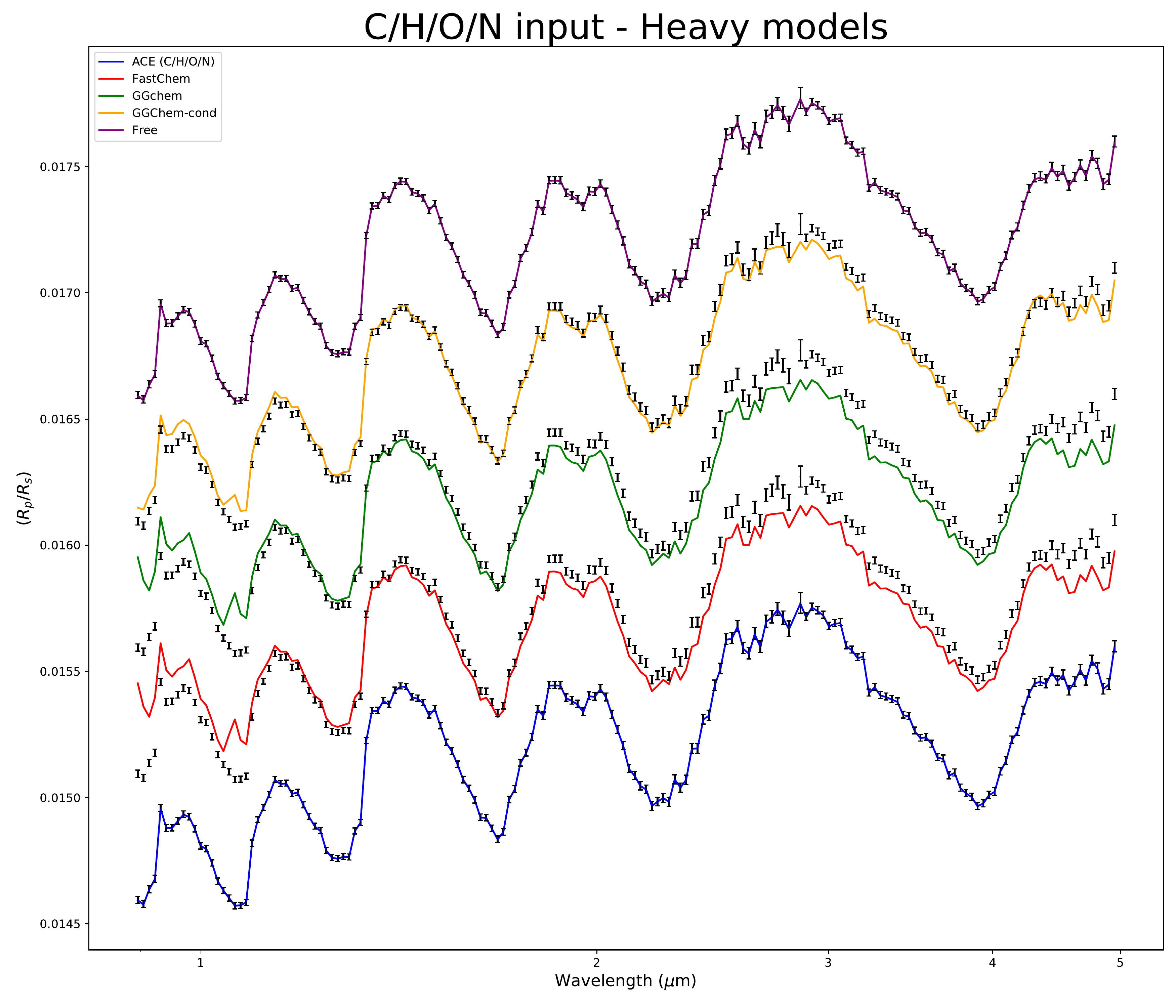}
\caption{Retrieved spectra for ACE (blue), FastChem (red), GGChem (green), GGChem with condensation (orange) and free chemistry (purple). Each equilibrium chemical code utilizes the heavy element list (except ACE and free) and is retrieved against simulated JWST spectra (black) computed with ACE using parameters from Table \ref{tab:hd209-test}. Spectra have been binned to a lower resolution and offset for clarity. }
\label{fig:chon-heavy-spec}
\end{figure}
\end{center}

We observed significant differences in the posteriors (Figure \ref{fig:chon-heavy-post}) when applying the heavy element list on FastChem and GGchem to the C/H/O/N spectrum. The heavy only element list have no direct mechanism to remove TiO and VO from the atmosphere due to our choice of priors, so the sampler attempts to mask their features underneath H$_2$O for both GGchem and Fastchem by increasing its abundance (Figure \ref{fig:chon-heavy-prof}) through the metallicity parameter up to the prior boundaries at 4.0$Z_{\odot}$. The increased metallicity introduces more strong carbon-bearing features. However, these are muted by a significant reduction in the C/O ratio to 0.15. The increase in depth by H$_2$O is compensated for through a reduction in the planetary radius. The resulting heavy spectra in Figure \ref{fig:chon-heavy-spec} are almost entirely water with a reduction in depth at 2\micron\ and 3--5\micron\ by the loss of CO and CH$_4$.  The heavy element list with condensation also employs a similar tactic but not as aggressively as it has the means to reduce gas-phase TiO and VO through condensation. The condensate mixing profiles in Figure \ref{fig:chon-heavy-prof} show a higher altitude of stability with an increase to 10$^{-5}$ bar for MgTiO$_3$ and 10$^{-8}$ bar for Ti$_4$O$_7$.

The resultant spectrum in Figure \ref{fig:chon-heavy-spec} for this case retains many of the spectral features seen in the C/H/O/N chemistry with a significant reduction in TiO and VO absorption features in the optical.

\subsubsection{Heavy input spectrum}

\begin{center}
\begin{figure}[ht]
    \includegraphics[width=1.0\columnwidth]{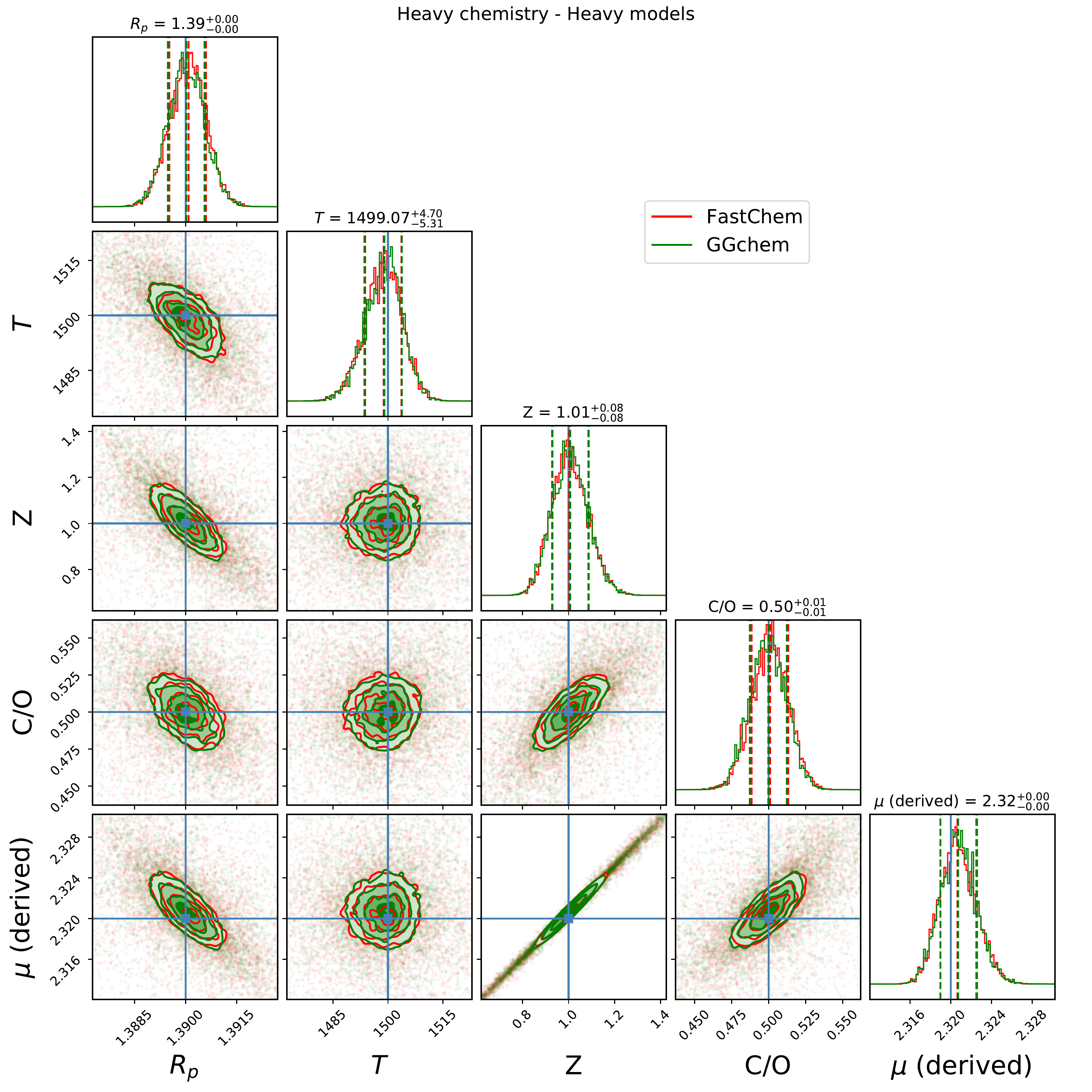}
\caption{Retrieval posteriors FastChem (red), GGChem (green) only. Each chemical code utilizes the heavy element list and is fit against a JWST simulated transit spectra of HD 209458b using parameters given by Table \ref{tab:hd209-test} and abundances computed using FastChem \citep{fastchem} with heavy species. Priors used are given by Table \ref{tab:chem-priors}.}
\label{fig:heavy-heavy-only}
\end{figure}
\end{center}

We examine the case where the retrievals fit against a heavy species spectrum. When examining only FastChem and GGChem, the posteriors in Figure \ref{fig:heavy-heavy-only} display a high degree of overlap and shape similarity. Both achieve almost the same correlations between parameters and tightly constrain their posteriors.

\begin{center}
\begin{figure}[ht]
    \includegraphics[width=1.0\columnwidth]{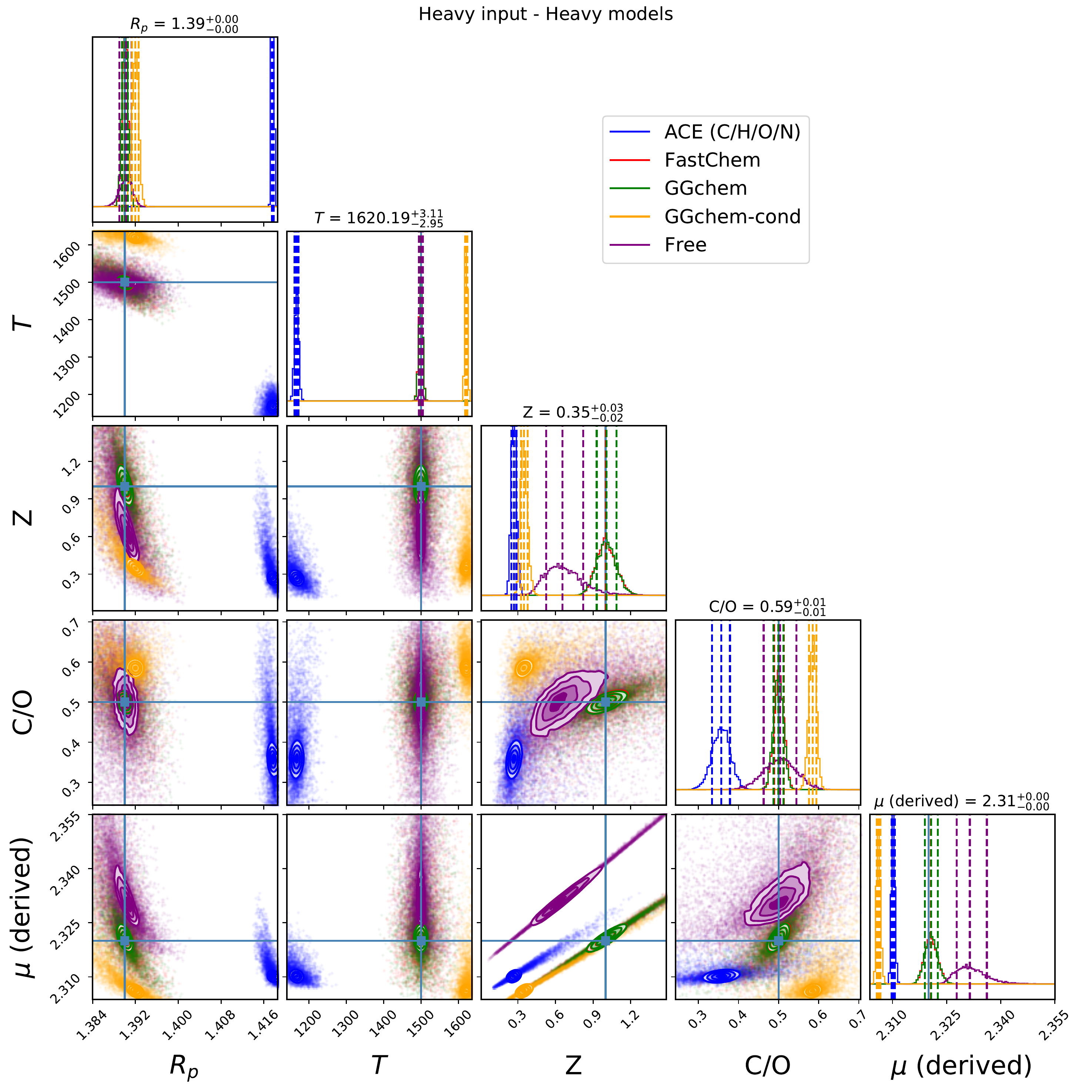}
\caption{Posteriors of ACE (blue), FastChem (red), GGChem (green), GGChem with condensation (orange) and free chemistry (purple). Each equilibrium chemical code utilizes the heavy element list (except ACE and free) and is fit against a JWST simulated transit spectra of HD 209458b using parameters given by Table \ref{tab:hd209-test} and abundances computed using FastChem\citep{fastchem} with heavy species. Priors used are given by Table \ref{tab:chem-priors}.}
\label{fig:heavy-heavy-post}
\end{figure}
\end{center}

\begin{center}
\begin{figure}[ht]
    \includegraphics[width=1.0\columnwidth]{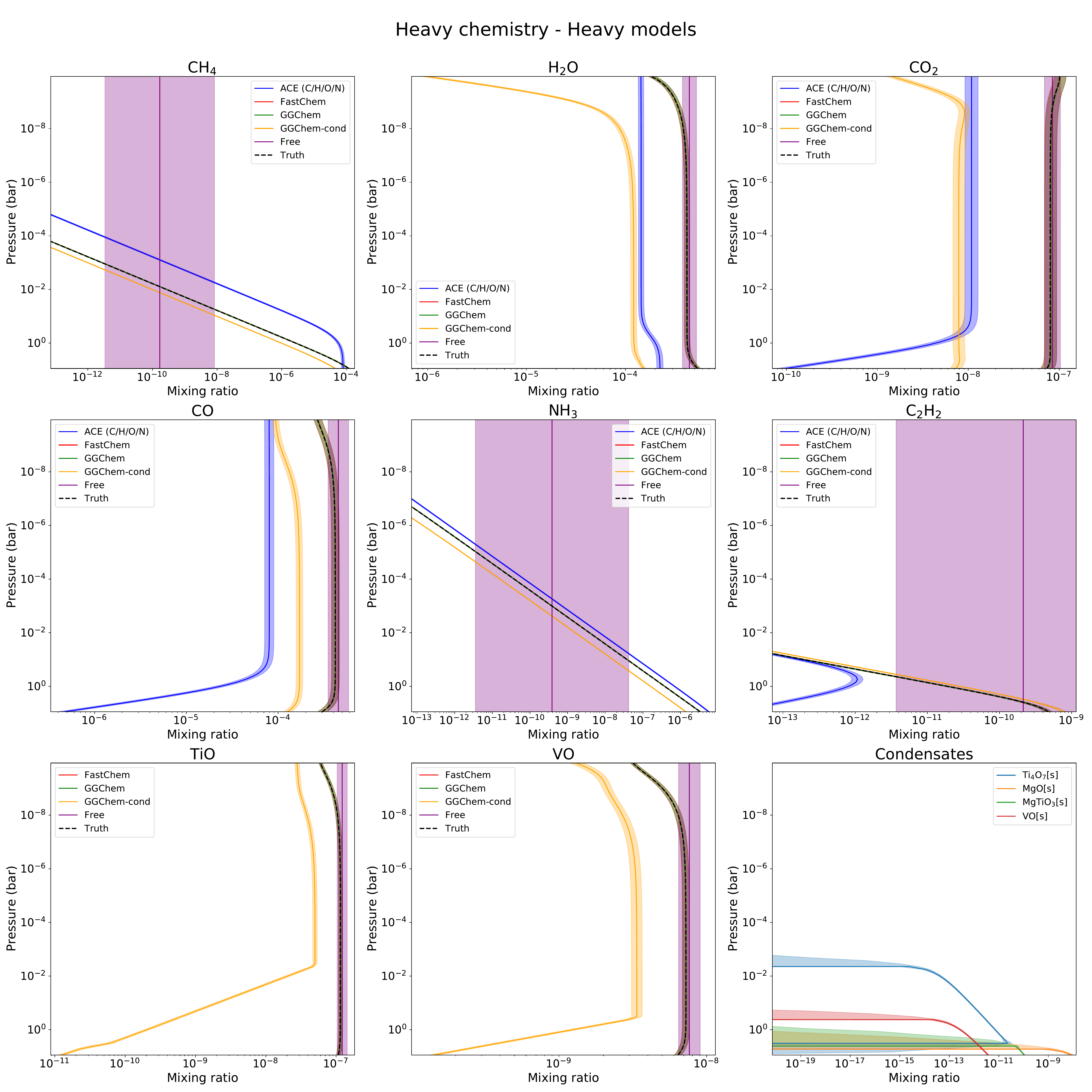}
\caption{Retrieved molecular profiles for ACE (blue), FastChem (red), GGChem (green), GGChem with condensation (orange) and free chemistry (purple). Each equilibrium chemical code utilizes the heavy element list (except ACE and free). Shaded regions denote 1$\sigma$ span, dashed lines are truth values computed from the GGChem chemical code with heavy species using Table \ref{tab:hd209-test}.}
\label{fig:heavy-heavy-prof}
\end{figure}
\end{center}

\begin{center}
\begin{figure}[ht]
    \includegraphics[width=1.0\columnwidth]{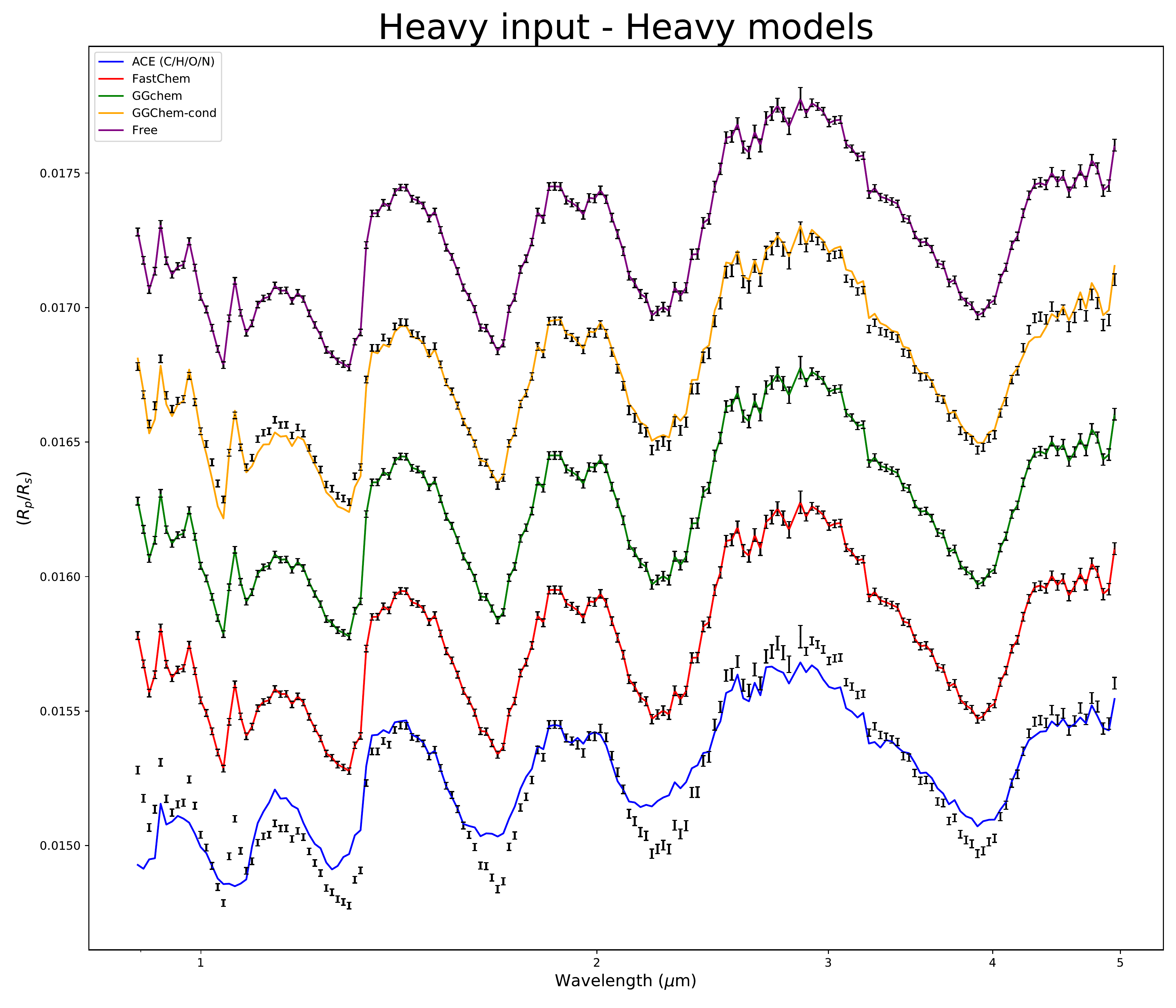}
\caption{Retrieved spectra for ACE (blue), FastChem (red), GGChem (green), GGChem with condensation (orange) and free chemistry (purple). Each equilibrium chemical code utilizes the heavy element list (except ACE and free) and is retrieved against simulated JWST spectra (black) computed with FastChem with heavy chemistry using parameters from Table \ref{tab:hd209-test}. Spectra have been binned to a lower resolution and offset for clarity.}
\label{fig:heavy-heavy-spec}
\end{figure}
\end{center}

Looking at the retrieval in its full context, the posteriors in Figure \ref{fig:heavy-heavy-post} include both ACE and GGchem with heavy condensation, and we see a significant deviation in their resultant posteriors. 
The resultant molecular profiles in Figure \ref{fig:heavy-heavy-prof} and spectra in Figure \ref{fig:heavy-heavy-spec} reveal the different approaches both ACE and GGchem with condensation take to reach their  retrievals. 
For the ACE case (and C/H/O/N chemistry in general), the sampler attempts to raise the features at 1\micron\ by increasing the amount of CH$_4$ in the atmosphere while lowering the contribution by CO.

This effect manifests by significantly reducing the amount of oxygen (i.e., the metallicity) as well as slightly reducing the C/O ratio. The knock-on effect of the reduction of oxygen lowers the overall transit depth as the abundance of H$_2$O drops significantly. The sampler counteracts this change by increasing the planetary radius to about 1.41 Jupiter radii. 
By contrast, the condensation retrieval attempts to minimize the condensates present to maximize the available TiO and VO. Looking at the condensation posteriors in Figure \ref{fig:heavy-heavy-post}, the higher temperature, and lower metallicity drop the stability and abundance of the condensates down to 10$^{-3}$ pressure range with some condensates like MgTiO$_3$ and MgO only appearing at pressures near the surface. The loss of oxygen is counteracted with a slight increase in the C/O ratio to maintain CH$_4$ abundance and planet radius to maintain relative depths. The resultant spectrum in Figure \ref{fig:heavy-heavy-spec} restores most of the spectral features present in the heavy only species with only slight losses in VO opacity near 1.2~\micron\ and CO at 4~\micron. Both Fastchem and GGchem retrieve almost exactly the same abundances for all molecules present.

Concerning the free chemistry models, the retrieved chemical profiles are well representing the inputs (from the heavy chemical models) for H$_2$O, CO, CO$_2$, VO and TiO, which are the most abundant molecules in the input spectrum. For CH$_4$, NH$_3$ and C$_2$H$_2$, the retrieval only provides a upper limit of about 10$^{-7.5}$ due to the lack of features of these molecules in the spectrum.

\subsubsection{Heavy with condensation input spectrum}

\begin{center}
\begin{figure}[ht]
    \includegraphics[width=1.0\columnwidth]{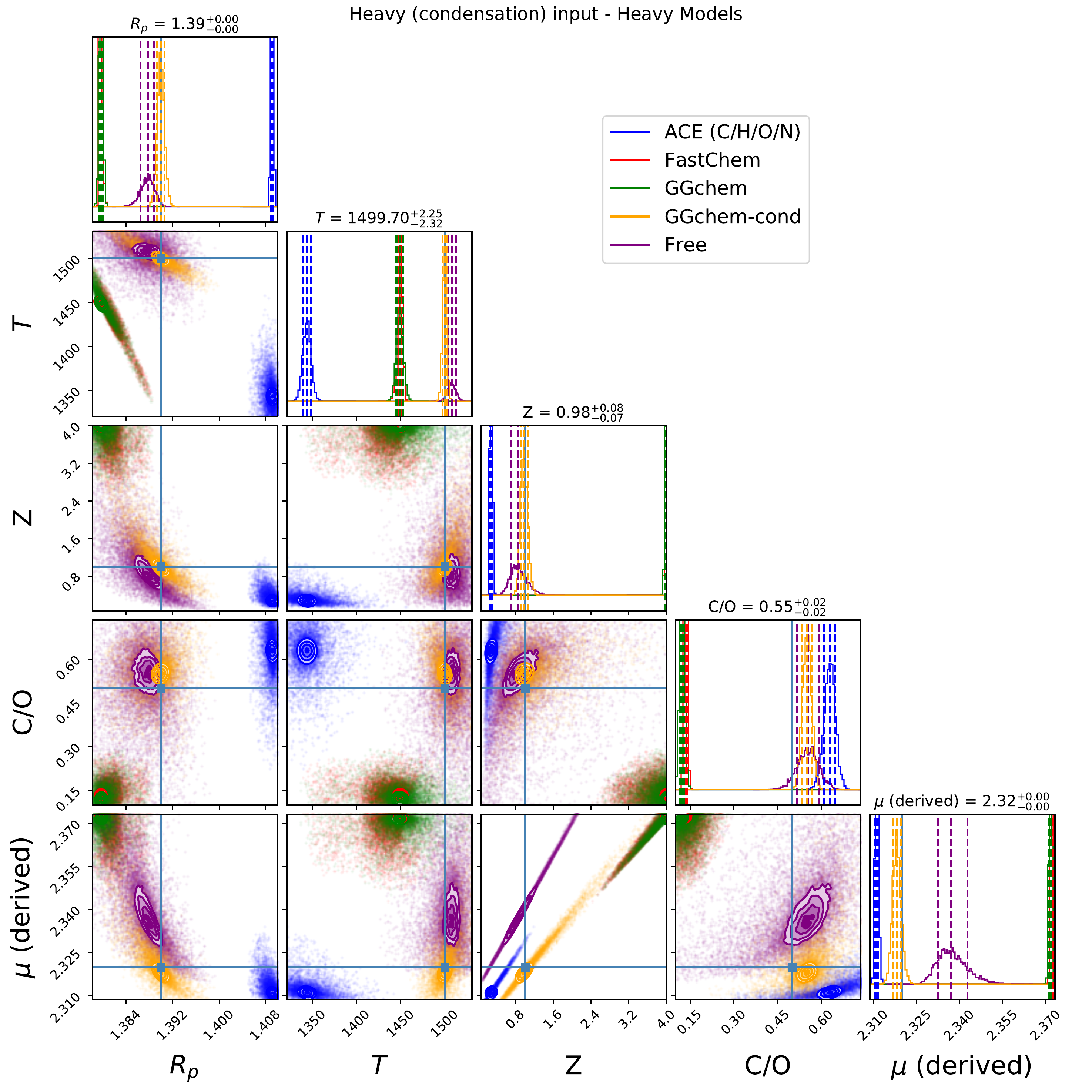}
\caption{ACE (blue), FastChem (red), GGChem (green), GGChem with condensation (orange) and free chemistry (purple). Each equilibrium chemical code utilizes the heavy element list (except ACE and free) and is fit against a JWST simulated transit spectra of HD 209458b using parameters given by Table \ref{tab:hd209-test} and abundances computed using GGChem\citep{fastchem} with heavy species and condensation. Priors used are given by Table \ref{tab:chem-priors}.}
\label{fig:heavy-cond-post}
\end{figure}
\end{center}

\begin{center}
\begin{figure}[ht]
    \includegraphics[width=1.0\columnwidth]{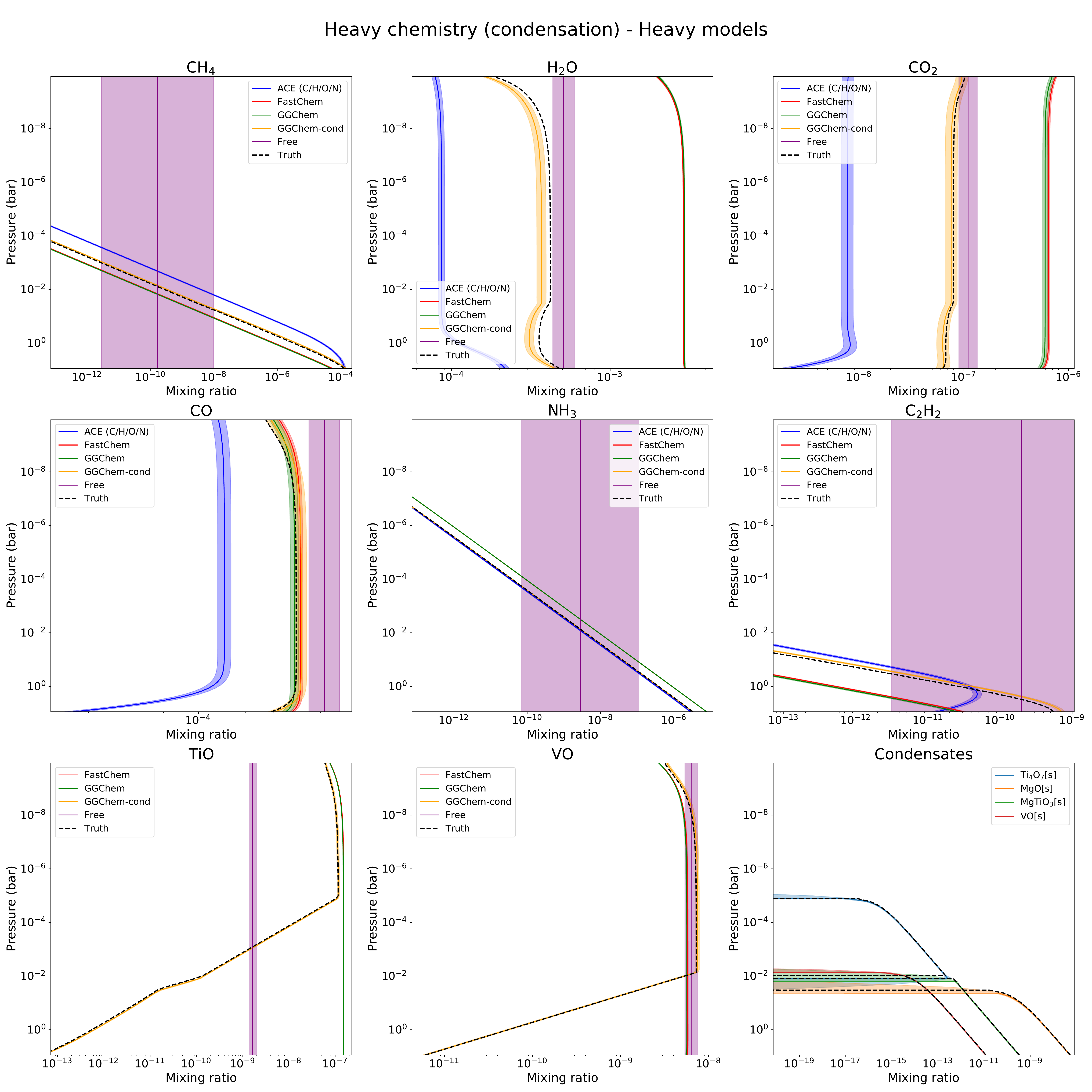}
\caption{Retrieved molecular profiles for ACE (blue), FastChem (red), GGChem (green), GGChem with condensation (orange) and free chemistry (purple). Each equilibrium chemical code utilizes the heavy element list (except ACE and free). Shaded regions denote 1$\sigma$ span, dashed lines are truth values computed from the GGChem chemical code with heavy species and condensation using Table \ref{tab:hd209-test}.}
\label{fig:heavy-cond-prof}
\end{figure}
\end{center}

We finally consider the case where all chemical codes retrieve against a heavy condensation truth. We will not consider C/H/O/N with condensation, as it is evident in Figure \ref{fig:chon-chon-post} that it is equivalent to C/H/O/N with no condensation for temperatures considered. For the C/H/O/N only species, we see similar behavior to the previous comparison in Figure \ref{fig:heavy-cond-post} as it is increasing the CH$_4$ abundance to raise the overall baseline spectrum. However, the lower TiO and VO signature results in the temperature dropping only to 1350~K compared to the 1120~K in the previous case and the reduction in metallicity only to 0.7 $Z_{\odot}$. Similarly, the heavy models mask TiO and VO by raising the water abundance, as seen in Figure \ref{fig:heavy-cond-prof}. Since there is still TiO and VO present, the temperature difference from truth is only 60~K compared to the 150~K drop in the C/H/O/N case. 
The results of the free retrieval on the heavy with condensation case are similar to the one from the heavy case, except for TiO and VO. Due to condensation, TiO and VO are sequestered from the bottom of the atmosphere, which leads the free constant retrieval to average these abundances. We believe a more complex parametric description of those profiles, such as the one presented in \cite{Changeat19layer}, would avoid those biases. Due to the biases arising from the constant assumption on TiO and VO, the H$_2$O, CO and CO$_2$ are more abundant in the atmosphere, causing a higher metallicity in the posteriors.

\begin{center}
\begin{figure}[ht]
    \includegraphics[width=1.0\columnwidth]{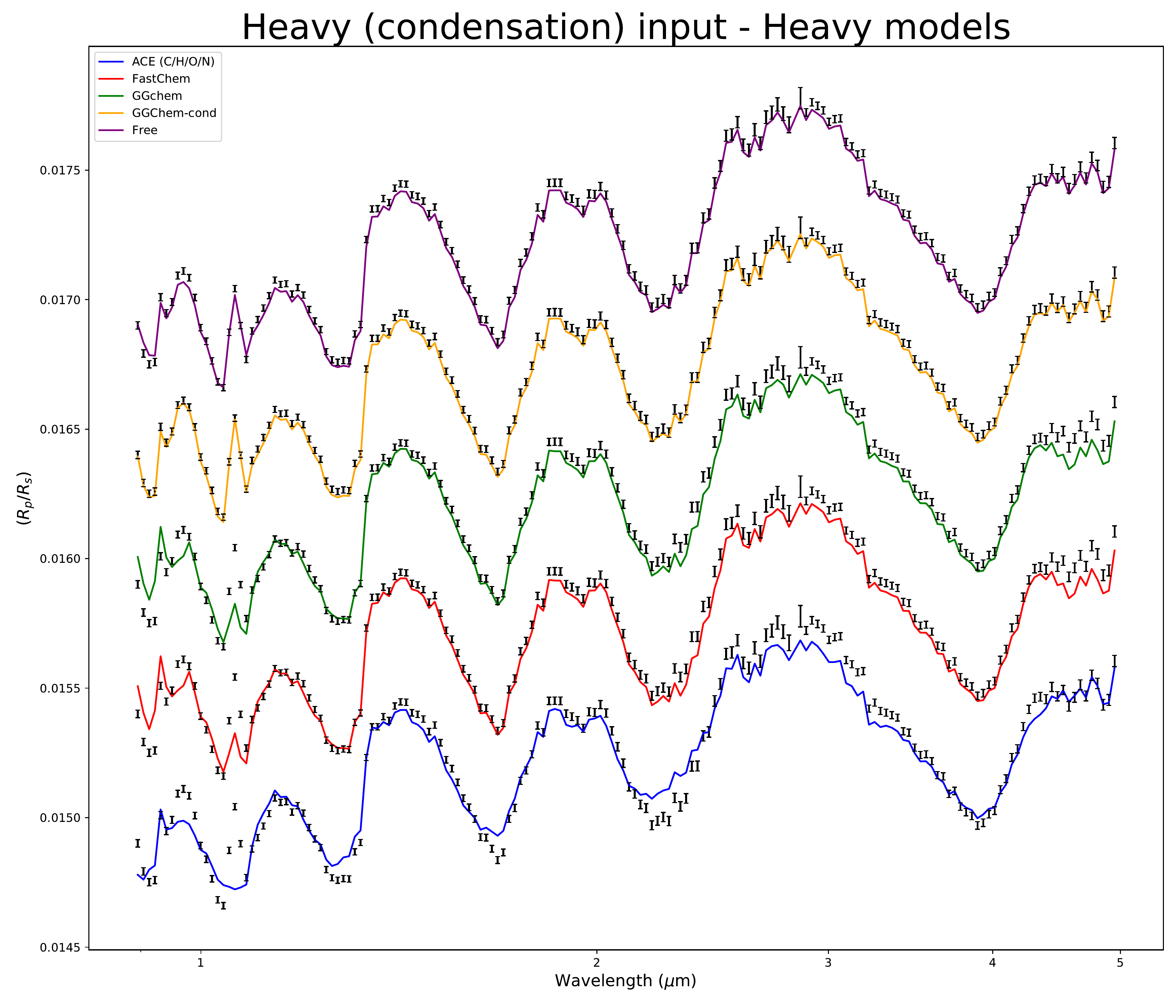}
\caption{Retrieved spectra for ACE (blue), FastChem (red), GGChem (green), GGChem with condensation (orange) and free chemistry (purple). Each equilibrium chemical code utilizes the heavy element list (except ACE and free) and is retrieved against simulated JWST spectra (black) computed with GGChem with heavy elements and condensation using parameters from Table \ref{tab:hd209-test}. Spectra have been binned to a lower resolution and offset for clarity}
\label{fig:heavy-cond-spec}
\end{figure}
\end{center}

Overall, the final spectra in Figure \ref{fig:heavy-cond-spec} show that the C/H/O/N and heavy case agree qualitatively better than their respective counterpart retrievals (see Figure \ref{fig:heavy-cond-post}).

\section{DISCUSSION}

When the model species and input spectrum species do not match, we find a significant deviation between true and retrieved values. For example, the retrieved uncertainties for temperature are less than 1\% even when the deviation from the truth is greater than 8\%. Far worse are the chemistry parameter values; When comparing C/H/O/N models with heavy input spectra, the metallicity posteriors have a minor deviation ($\pm$ 0.02) than models with the same element list ($\pm 0.08$) even though the retrieved value deviates by over 20\%. Comparably the uncertainties for the retrieved C/O ratio also exhibit the same behaviour while deviating from the truth by 10\%. Similarly, introducing condensation into the input spectrum appears to retrieve metallicity posteriors with significantly smaller variances and can erroneously attribute the planet to having subsolar composition. Finally, comparing heavy chemistry with C/H/O/N atmospheres, we find the retrieval is incapable of building posteriors within the tight prior bounds for metallicity and C/O ratio, with the retrieval moving towards a high metallicity value. These retrievals demonstrate that the assumptions we make on the composition and processes (i.e. choice of elements, condensation etc.) occurring in the atmosphere are the significant source of bias. Unfortunately, we have no prior knowledge of the underlying physics within exoplanetary atmospheres in a realistic setting, so immediately assuming a self-consistent model can lead to confident and incorrect conclusions of atmospheric composition.

In the context of the next generation, extensive surveys \cite[Ariel: ][]{Tinetti_2021_ariel} and dedicated studies \cite[JWST: ][]{Greene_2016} of exoplanets, the higher information content of spectra observed means that biases from our assumptions becomes significantly more pronounced than in HST data and greater care must be taken in analysing exoplanetary atmosphere. Using less complex models and assumptions such as isoabundance initially will allow for careful exploration of the information content of the spectra and identification of species and processes. Following this approach would allow us to make an informed constraint on priors when introducing more complexity and self-consistency.
Next-generation retrievals must have the flexibility to include a wide range of processes and increase/decrease its model complexity to interpret spectra successfully.



\section{CONCLUSIONS}

We have presented here  one of the first systematic comparisons of equilibrium chemical codes in the context of exoplanet retrievals. This comparison was made
possible by the newest version of TauREx 3.1, allowing each chemical code to be `plugged-in' to the framework. TauREx 3.1 and its available plugins are available on PyPI as source distributions and binaries for Windows, macOS and Linux. The sources for TauREX 3.1 and its plugins are also available on the ucl-exoplanet GitHub\footnote{\url{https://github.com/ucl-exoplanets}} with permissible licenses on most plugins. We hope that providing a framework with a plugin interface will allow for more collaboration and interconnectivity between different fields and future comparative work.
Regarding the chemical comparisons, we demonstrate that ACE, GGchem and FastChem reach the same conclusions given the same assumptions. The implementation, source of thermochemical data and fitting of equilibrium functions introduce little to no bias in retrievals. The most significant source of bias comes from  assumptions on the chemical processes in the atmosphere and that these retrievals arrive at confident but incorrect solutions. To overcome such biases, free retrievals of the chemistry should be considered as a first step. This is particularly relevant for next generation of telescopes, such as JWST and Ariel, which will be sensitive to thermal chemical and dynamical processes that are not yet fully understood. Until our understanding chemistry in exoplanets can evolve, simpler approaches relying on the information content of the observations have to be favored.\\


\section*{Acknowledgements}
This project has received funding from the European Research Council (ERC) under the European Union's Horizon 2020 research and innovation programme (grant agreement No 758892, ExoAI) and the European Union's Horizon 2020 COMPET programme (grant agreement No 776403, ExoplANETS A). Furthermore, we acknowledge funding by the UK Space Agency and Science and Technology Funding Council (STFC) grants: ST/K502406/1, ST/P000282/1, ST/P002153/1, ST/S002634/1, ST/T001836/1 and ST/V003380/1. The authors would like to acknowledge P. Woitke, C. Helling, D. Kitzmann and J. Stock for providing their codes under permissive open-source. AFA would also like to thank Quratulain Jahangir for her support. OV acknowledges the CNRS/INSU Programme National de Plan\'etologie (PNP) and CNES for funding support.

This work utilised the OzSTAR national facility at Swinburne University of Technology. The OzSTAR program receives funding in part from the Astronomy National Collaborative Research Infrastructure Strategy (NCRIS) allocation provided by the Australian Government. This work utilised the Cambridge Service for Data Driven Discovery (CSD3), part of which is operated by the University of Cambridge Research Computing on behalf of the STFC DiRAC HPC Facility (www.dirac.ac.uk). The DiRAC component of CSD3 was funded by BEIS capital funding via STFC capital grants ST/P002307/1 and ST/R002452/1 and STFC operations grant ST/R00689X/1. DiRAC is part of the National e-Infrastructure.

\appendix

\section{\trex 3.1}
\label{ap:t31}
\trex{3.1} is the next version of the \trex{3} library, backward-compatible with the previous versions input files but offering a swathe of improvements and optimizations to the overall architecture.
The goal of this version was to expand the dynamic architecture and provide significantly more flexibility to the \taurex\ framework.



\subsection{Non-uniform priors}

We include a new type of class into the retrieval framework that handles the prior transform in Bayesian retrievals. The \texttt{Prior} class allows for the inclusion of custom functions to be used as the prior transform and removes the uniform-only limitation of the previous version. A new \textit{prior} flag can be used when in the input file definition of the fitting parameters: 
\begin{Verbatim}[frame=single,fontsize=\small]
[Fitting]
planet_radius:fit = True
planet_radius:prior = "LogUniform(bounds=[-2,2])"
\end{Verbatim}
This is equivalent to the previous versions definition of fitting priors:
\begin{Verbatim}[frame=single,fontsize=\small]
[Fitting]
planet_radius:fit = True
planet_radius:mode= log
planet_radius:bounds=1e-2,1e2
\end{Verbatim}
It must be noted that the the older input definition is still compatible with \trex{3.1}, however, internally they are converted into the new \texttt{Prior} form.
On installation there are only 4 functions available: \texttt{Uniform}, \texttt{LogUniform}, \texttt{Gaussian} and \texttt{LogGaussian}. Installing the \texttt{taurex\_scipypriors} plugins increases to over 50 different functions.

\subsection{H- opacity} \label{sec:Hminus}

A new \texttt{HydrogenIon} contribution class can be included in both transmission and emission forward models that provide opacities from continuous absorption of H$^{-}$. We compute the absorption opacity using equations 3--6 from \cite{John88} for free-free and bound-free transitions and require the chemistry model to provide atmospheric abundances of gas-phase $H$ and $e-$. The algorithm is optimized heavily, requiring only 500$\mu$s to run for R=15,000.

\section{Plugins}
\label{ap:plugins}
A key feature in the first release of \trex{3} allowed the user to inject their code into the retrieval pipeline from the input file. 
This feature allowed for the inclusion of new atmospheric models and parameters in the retrievals without extensive knowledge of the underlying systems within \trex{3}. 
However, these codes are considered exceptions from the standard \trex{3} pipeline and required specific input keywords and files in order to function. 
This way also prevents the accumulation of enhancements to the framework provided from the feature as they must now obtain a specific code file. 
The distribution of such custom enhancements was cumbersome, especially for FORTRAN and C++ codes, as they would require additional manual stages before they could be used. 
In most cases, developers directly placed their code into their copy of the \trex{3} codebase. 
This approach, while valid, runs the risk of splintering \trex{3} into multiple, incompatible versions.
The latest version of \trex{3} (3.1+) remedies this problem through the use of \textit{plugins}. Plugins are, in essence, installable enhancements to \trex{3}. 
They allow new features to be included in the pipeline natively without modifying the main \trex{3} codebase. 
A plugin author can build new chemistries, profiles, models and optimizers and distribute them to other users through the Python packaging system.

To give an example, when \trex{3.1} is installed (or upgraded from version 3.0), only \texttt{free}
chemistry is available. 
An arbitrary input file for that may look like this:

\begin{Verbatim}[frame=single,fontsize=\small]
[Chemistry]
chemistry_type = free
fill_gases = H2,He
ratio = 0.17
    [[H2O]]
    gas_type = constant
    mix_ratio = 1e-3
    [[NH3]]
    gas_type = constant
    mix_ratio = 1e-5

[Fitting]
H2O:fit = True
H2O:priors = "LogGaussian(mean=-1,std=0.25)"
\end{Verbatim}

No other chemical model is available. However, if an equilibrium model such as GGChem \citep{ggchem} is desired, the user can run:
\begin{minted}{bash}
$ pip install taurex_ggchem
\end{minted}

At which point, the plugin is automatically downloaded, compiled if needed (or prebuilt binaries used) and installed. Now GGChem becomes available and can be utilized in the input file with its own set of input keywords and retrieval parameters:
\begin{Verbatim}[frame=single,fontsize=\small]
[Chemistry]
chemistry_type = ggchem
equilibrium_condensation = True
metallicity = 1.0
selected_elements = H, He, C, N, O
new_back_it = 6
include_charge = False

[Fitting]
C_O_ratio:fit = True
C_O_ratio:priors = "Uniform(bounds=[1e-2, 2.0])"
\end{Verbatim}

On initialization, \trex{3} searches for modules in the python environment for these plugins and adds any new keywords to be parsed in the input file. \trex{3} does not assume the inclusion of its inbuilt classes and parameters and will perform the same search of parameters and models on itself. The plugins function like any other Python library can be installed from PyPi and git through \texttt{pip install} which significantly
improves their ability to be distributed to other users.
This ability to be installed also allows for the compilation stages of FORTRAN and C++ codes to be completely automated and transparent to the user, significantly improving their usability. Importantly, the original author has full control of development, able to host and develop their own plugins independently of \trex{3}. The plugins for our source control forks of the FastChem \citep{fastchem} and GGchem equilibrium chemistries introduce minor or no changes to the original code. Instead, the installation stage references and compiles against the chemistry author's original FORTRAN/C++ code. Once the TauREx Python wrappers and plugin setup files have been written, an author can continue development in their language of choice; Leaving their original makefiles and overall compilation pipeline intact and allowing the plugin to benefit from their newest improvements. 
Development of these wrappers is also accessible. A developer can only focus on converting TauREx inputs and units to their desired input, performing the calculation, and converting them into the TauREx expected outputs and units. As TauREx standardizes each atmospheric component, there is a guarantee that the implementation will fully function in the retrieval pipeline.

Plugins also solve another problem of distribution of codes. Some of these codes may have restrictive licenses that
prevent them from being fully open-source. These plugins may instead be installed directly from private repositories or distributed as compiled python code (i.e \texttt{.pyc} instead of \texttt{.py}) and binaries with licenses.
A list of the available plugins are given in Table \ref{tab:plugin-list}.
\begin{table}[ht]

\centering
\begin{tabular}{llr}
\hline\hline
Plugin & Description & Availability  \\
\hline
\texttt{taurex\_cuda} & CUDA-acceleration of forward models  & PyPI\\
\texttt{taurex\_hip} & HIP-acceleration of forward models & PyPI \\
\texttt{taurex\_ace} & Equilbrium Chemistry using ACE & PyPI \\ 
\texttt{taurex\_ggchem} & Equilbrium Chemistry using GGChem & PyPI \\
\texttt{taurex\_fastchem} & Equilbrium Chemistry using FastChem & PyPI\\
\texttt{taurex\_ultranest} & Ultranest sampler support for TauREx & PyPI\\
\texttt{taurex\_dynesty} & Dynesty sampler support for TauREx & PyPI \\
\texttt{taurex\_scipypriors} & \texttt{scipy.stat} continuous functions as priors & PyPI \\
\texttt{taurex\_petitrad} & petitRADTRANS forward models and opacity formats & PyPi \\
\texttt{taurex\_catalogue} & Set planetary and stellar parameters from name & On publication\\
\texttt{taurex\_uv} & UV stellar spectra slicing & On publication\\
\texttt{taurex\_phasecurve} & 1.5D phasecurve forward models & On publication \\
\texttt{taurex\_jwst} & JWST instrument noise simulator & On publication \\

\hline
\end{tabular}
\caption{List of currently available plugins of \trex{3.1}. Availability describes where a user may acquire the plugin. \textit{PyPI} availability means that the plugin can be acquired using \texttt{pip install}. \textit{On publication} refers to plugins that will be available after their relevant publication.}
\label{tab:plugin-list}
\end{table}

\subsection{TauREx-CUDA}
\label{sec:cuda}
One of the first plugins developed provides GPU acceleration to the forward models and retrievals using the \texttt{PyCUDA}
library. It is installed by executing \texttt{pip install taurex\_cuda} and provides replacement models and contributions functions that take advantage of nVidia GPU cards.
Once installed, utilizing the GPU requires replacing the forward models and contributions with \texttt{cuda} suffix versions in the input file. For example, a transmission model with Rayleigh scattering:
\begin{minted}{python}
[Model]
model_type = transmission
       [[Absorption]]
       [[Rayleigh]
\end{minted}
can be GPU enabled by replacing with TauREx-CUDA versions of the model and contributions:
\begin{minted}{python}
[Model]
model_type = transmission-cuda
       [[AbsorptionCUDA]]
       [[RayleighCUDA]
\end{minted}
The plugin also contains a cross-section caching system that will move
and reuse absorption and collisionally induced absorption (CIA) cross-sections within GPU memory.
The GPU opacities work slightly differently to the CPU counterpart as they now perform interpolation and weighting of all layers in the atmosphere in parallel. 
Similarly the integration of the atmospheric layer works the same way where
rather than calculating layer by layer, the contributions now parallel compute all wavelengths and
layers in the calculation.
All CUDA kernels within the plugin are generated on the fly, compiled and cached for later reuse.
This approach allows for the generation of optimized CUDA code for a particular calculation. For example, on the first run,
the calculation of the eclipse depth unwraps and inserts the values of the Gaussian-Quadrature weights and abscissa 
into the kernel source code. This completely eliminates global reads, significantly boosting performance. Table \ref{tab:cuda-gauss} highlights this approach as 50 point quadrature integration results in only a 40\% increase in modelling time against a 4 point quadrature on a V-100 GPU compared to a 4x increase for the CPU version.
\begin{table}[ht]
\centering
\begin{tabular}{lrrr}
\hline\hline
N$_{quads}$ & CPU (ms) & GPU (ms) & Speedup \\
\hline
4 & 1170 & 21.2 & 55\\
8 & 1280 & 23.6 & 54  \\
16 & 1500 & 24.1 & 62 \\ 
32 & 2140 & 33.4 & 64 \\
50 & 4560 & 37.2 & 122 \\
\hline
\end{tabular}
\caption{CPU and GPU computation time for an emission model using a varying number of Gaussian quadrature points. The atmosphere contains 100 layers with 5 actively absorbing molecules, CIA and Rayleigh scattering.  }
\label{tab:cuda-gauss}
\end{table}
This is important to consider as previous studies \citep{quentin_phase} have shown that a minimum quadrature of 10 points is needed to achieve relatively good convergence.

Assessing performance, Table \ref{tab:cuda-perf-r} shows the runtime of both transmission and emission models for a range of resolutions. The benchmarks were conducted on an Intel(R) Xeon(R) Gold 6130 CPU at 2.10GHz with a single 16GB nVidia V100 GPU. 
The performance gain from the CUDA accelerated contribution functions is significant with up to 150$\times$ reduction in modelling time compared to the CPU implementation. 
For the line-by-line case (R=1,000,000), memory pressure required limiting the benchmark to a single absorbing molecule. 
However, comparing like-for-like with the CPU version demonstrates a degree of viability in using line-by-line in retrievals in the transmission case. 
The current version of the CUDA plugin uses a lazy but straightforward memory management scheme. Cross-sections are cached in their entirety in GPU memory. The results of their interpolation and the optical depth itself are also generated in GPU memory. In the line-by-line scheme, each of these arrays occupies roughly 4 gigabytes of memory. The transmission case only requires 12 gigabytes to complete, which just about fits in a 16GB V100 GPU. The emission case requires two additional arrays which bring the memory cost to 20 GBs. An easy solution would be to use the 32GB variant of the V100 or exploit the newer 40GB A100 cards. Future versions of the plugin will aim to introduce a smarter memory management system that would significantly reduce memory usage through the exploitation of asynchronous memory transfers. With either the introduction of higher memory GPUs or smarter memory algorithms, we hope to transition to line-by-line retrievals shortly. 
\begin{table}[ht]

\centering
\begin{tabular}{lrrrr}
\hline\hline
 & Transmission & & Emission & \\
R & CPU (ms) & GPU (ms)  & CPU (ms) & GPU (ms)  \\
\hline
7,000      & 1319    & 10      & 1777 & 18     \\
10,000     & 1864    & 12      & 2758 & 12   \\
15,000     & 3055    & 16      & 4501 & 15    \\ 
1000000* & 121580 & 406    &   63040   & n/a   \\
\hline
\end{tabular}
\caption{CPU and GPU computation time for transmission and 4-point quadrature emission model for a range of resolutions at the full wavelength range of 0.3\micron--15\micron . The atmosphere contains 100 layers with 5 actively absorbing molecules, CIA and Rayleigh scattering. The benchmarks were conducted on an Intel(R) Xeon(R) Gold 6130 CPU at 2.10GHz with a single 16GB nVidia V100 GPU. Timing was conducted using the \texttt{timeit} python module.
*Due to memory constraints of the GPU the $R=1,000,000$ benchmarks were conducted without CIA and Rayleigh scattering and with only a single molecule. The source of the super high resolution opacity is from the line-by-line data from petitRADTRANS \citep{Mollire_petitrad} loaded using the \texttt{taurex\_petitrad} plugin. It has a wavelength range of 0.1\micron--28\micron.  }
\label{tab:cuda-perf-r}
\end{table}

All contribution functions are supported, including non-CUDA enabled ones. TauREx-CUDA will perform any non-CUDA
calculations first before transferring the results to the GPU and completing any remaining calculations.
Finally, the TauREx-CUDA plugin can be used to build new CUDA enabled contribution functions and models and also provides
a GPU opacity caching system for accelerated interpolation of molecular cross-sections.

\section{Chemistry Plugins}
\label{ap:chemistry}
The plugin system has now expanded the scope of chemistry modelling to include three equilibrium chemistries: ACE \citep{Agundez2012, Agundez20}, FastChem \citep{fastchem} and GGChem \citep{ggchem}. All plugins can be installed
through PyPi on Windows, Linux and Mac and all are capable of being used in optimizations to retrieve elemental abundances. ACE and GGchem are all originally written in FORTRAN. To facilitate
their inclusion, the numpy f2py module was used to automatically generate efficient and convenient python wrappers as well as handle their compilation during installation. FastChem is a C++ code so light wrappers
were written using cython. For all of these codes, a dedicated \trex{3} chemistry interface was written
and exposed to the plugin system.
When installed through PyPI, the python packaging system handles all of the wrapping and compilation in the background. For Windows and MacOS, PyPI will bypass compilation and instead download prebuilt binaries.

All chemistries present can be controlled and fit based on metallicity relative to their initial abundance (usually solar) and ratios of
each metal element relative to oxygen. For the ACE plugin, only C/O and N/O ratios can be fit. The Fastchem and GGchem plugins will dynamically generate oxygen ratio fitting parameters \citep{alrefaie2019taurex} for each metal element selected by the user. For example, if the user selects C, N, S and Ti then C/O, N/O, S/O and Ti/O ratios can be fit during retrievals.

\subsection{GGchem}

The plugin compiles against the original GGchem FORTRAN source code during installation as well as introducing additional FORTRAN-90 glue code that directly interfaces with \trex{3}. 
The GGchem plugin allows for the user selection of elements. Hydrogen, helium and oxygen must be present. If the user does not specify these then they are automatically included. The initial abundance of these elements can be chosen from one of four default profiles: \texttt{solar}, \texttt{earth}, \texttt{ocean} or \texttt{meteorite}.
Modification of abundances after initialization (and during retrievals) is controlled by the \texttt{metallicity} parameter which determines
the amount of oxygen relative to the initial profile and an \texttt{O\_ratio} parameter that determines the ratio for each metal element relative to oxygen. The oxygen ratio fitting parameters are dynamically generated \citep{alrefaie2019taurex}. The plugin also allows for equilibrium condensation, enabled at initialization by setting
\texttt{equilibrium\_condensation=True} and ions enabled by setting \texttt{include\_charge=True}. 
As ion chemistry defines electron abundance, it can be seamlessly used with the H- opacity component (Section \ref{sec:Hminus}).

The \texttt{taurex\_ggchem} plugin, to the authors knowledge, presents the only true python wrapper to GGchem.
All functionality of GGchem is present and keywords in the \trex{3.1} input file match (or closely match) a GGchem input file.
The plugin can be used outside of \trex{3} to generate chemistry profiles through a simple python interface:
\begin{minted}{python}
>>> from taurex_ggchem import GGChem
>>> gg = GGChem(metallicity=1.0,  
         selected_elements=['H','He','C','O','N','K'], 
         abundance_profile='earthcrust', 
         equilibrium_condensation=True) 
>>> nlayers = 10
>>> temperature = np.linspace(300,100,nlayers)
>>> pressure = np.logspace(5,-3, nlayers) # Pa
>>> gg.initialize_chemistry(nlayers,temperature,pressure)
>>> gg.gases
['H', 'He', 'C', 'O', 'N',..., 'N3', 'O3', 'C3H']
>>> gg.mixProfile
array([[4.75989782e-04, 4.93144149e-04, 5.10561665e-04, ...,
        2.89575385e-05, 2.47386006e-05, 2.10241059e-05],
       ...,
       [2.49670621e-16, 1.44224904e-16, 8.29805526e-17, ...,
        9.48249338e-42, 4.75884162e-42, 2.37999459e-42]])
>>> gg.condensates
['C[s]', 'H2O[s]', 'H2O[l]', 'NH3[s]', 'CH4[s]', 'CO[s]', 'CO2[s]']
>>> gg.condensateMixProfile
array([[0.00000000e+00, 0.00000000e+00, 0.00000000e+00,...,
        0.00000000e+00, 0.00000000e+00],
       [0.00000000e+00, 0.00000000e+00, 0.00000000e+00, ...,
        0.00000000e+00, 9.82922802e-10, 1.88551848e-10, 2.88471985e-11,
        4.40651877e-12, 6.95597887e-13],
        ...,
        [0.00000000e+00, 0.00000000e+00, 0.00000000e+00, ...,
        0.00000000e+00, 0.00000000e+00]])
\end{minted}

GGChem presents certain features of FORTRAN that
make integration with Python more difficult. Firstly it possesses a global state meaning only one instance
of GGChem can execute in a single python script. Attempting to create two seperate GGchem instances will either crash
the code or generate incorrect results as the global state is being shared and modified between instances. 
An example of this would be if a user is attempting to generate forward models for multiple planets. The second issue comes from FORTRAN's \texttt{STOP} command which at its lowest level is a system call that kills the running process. 



Retrievals that encounter this halt and crash TauREx immediately. For GGchem, this is triggered when it cannot converge to a result. Rather than stopping completely,
it would be more desirable to continue a retrieval while avoiding these regions.
The simplest solution would be to refactor code to remove these issues but this can be a major undertaking
for authors of these codes and inherently goes against the philosophy of the plugin system. An alternative solution is to use the \texttt{SafeFortranCaller} class introduced in \trex{3.1} which circumvents these issues.
The class utilizes the python \texttt{multiprocessing} module to spawn a child python process that will execute the FORTRAN code. User requests for reading and writing variables or subroutine calls are passed as messages to the child process which will return the results back to the parent process.
With this method, the global state is now local to the child process. If another GGchem instance is required, a new child process is spawned. During a get/set/call request, the parent process will monitor the child process,
if at some point the child process has died, it is assumed that a \texttt{STOP} command was called. The parent process will
perform cleanup of hanging threads and raise a \texttt{FortranStopException}. On the next get/set/call request,
a new process is spun up. An additional benefit to this class is that all writes to standard output in the FORTRAN code are
also redirected back to \trex{3} and logged. This is useful as FORTRAN outputs can now be hidden during retrieval.

The GGchem plugin utilizes this class extensively. Multiple instances can now be created and retrievals freely
run. Anytime a convergence issue is encountered, the instance is destroyed, GGchem reinitialized and an exception
raised. The retrieval is informed of this and will avoid regions in the parameter space that cause it.

\subsection{ACE}
Previously the ACE equilibrium chemistry was built in to the original \trex{3}. For the newest
release it has been removed and turned into a plugin. This was to simplify the installation 
process of the main \trex{3} code and removed the need for a FORTRAN compiler to be present
in the users computer. With this, \trex{3} has a full Python stack. To restore the previous functionality,
a user need only install the \texttt{taurex\_ace} plugin in order utilize ACE as before.

\subsection{FastChem}

The FastChem plugin, like the GGChem plugin, allows for the calculation of ion chemistry, can be used outside of \trex{3} to generate chemistry profiles using Python:
\begin{minted}{python}
>>> from taurex_fastchem import FastChem
>>> fc = FastChem(selected_elements=['H','He','C','O','N','K','e-'], 
                  with_ions=True, metallicity=1.0)
>>> nlayers = 10
>>> temperature = np.linspace(300,100,nlayers)
>>> pressure = np.logspace(5,-3, nlayers) # Pa
>>> fc.initialize_chemistry(nlayers,temperature,pressure)
>>> fc.gases
['H', 'He', 'O', 'C', 'K', 'N', 'e-', ..., 'O+', 'O-', 'O2+', 'O2-']
>>> fc.mixProfile
array([[3.87435866e-036, 9.95149979e-039, 7.62616463e-042,
        1.23490910e-045, 2.58839801e-050, 3.41640407e-056,
        9.40930967e-064, 9.08433703e-074, 1.41255491e-087,
        1.38065040e-167],
        ...,
       [1.42400626e-001, 1.42400626e-001, 1.42400626e-001,
        1.42400626e-001, 1.42400626e-001, 1.42400791e-001,
        1.42398731e-001, 1.42398284e-001, 1.42367067e-001,
        9.96186945e-001]])
\end{minted}

Under the hood, the FastChem plugin generates parameter and custom elemental abundance files in a temporary directory that is then passed into the FastChem library. Doing this allows us to vary the elemental abundances through the oxygen ratio and metallicity parameters. Like GGChem, the retrieval parameters include metallicity which controls the amount of oxygen relative to the initial abundance and an \texttt{O\_ratio} parameter that determines the ratio for each selected metal element relative to oxygen. The initial abundance can be modified by passing in FastChem abundance files through the \texttt{elements\_abundance\_file} keyword. Once more, enabling ions, FastChem computes electron abundances which can be used be the H- opacity code (Section \ref{sec:Hminus}).

\bibliography{references}

\end{document}